\documentclass[reprint, showpacs, amsmath, amssymb, amsfonts, prd, nofootinbib, showkeys]{revtex4-1}
\usepackage[utf8]{inputenc}
\usepackage[T1]{fontenc}
\usepackage[ngerman,english]{babel}
\usepackage{graphicx}
\usepackage[dvipsnames, svgnames, table]{xcolor}
\usepackage{siunitx}
\usepackage[matha]{mathabx} % To use better looking arrows
\usepackage{multirow}
\usepackage{booktabs,lipsum,rotating,calc}
\usepackage{subfigure}      % to have subfigures
\renewcommand{\thesubfigure}{\alph{subfigure}}
\makeatletter
  \renewcommand{\@thesubfigure}{(\thesubfigure)\space}
  \def\@currentlabel{\p@subfigure\thesubfigure}
\makeatother
\usepackage{capt-of} %for the \captionof command
\usepackage{tikz, moresize, pgfplotstable}
\usetikzlibrary{positioning, calc, decorations.pathreplacing}
\pgfplotsset{compat=1.14}
\usepackage{hyperref}
\hypersetup{
    colorlinks=true,
    citecolor=black,
    linkcolor=black,
    urlcolor=black,
    anchorcolor=black,
    linktocpage % this links only the page numbers
}
\usepackage{etoolbox} % for \appto
\usepackage{cleveref}
\makeatletter%   https://tex.stackexchange.com/a/266344 needs etoolbox
\appto{\appendix}{%
  \@ifstar{\def\theequation@prefix{A.}}%
          {}%
}
\makeatother
\graphicspath{{./Figures/}}
\newcommand{\tc}[2]{\textcolor{#1}{#2}}
\newcommand{\referencename}{Ref.}
\newcommand{\refcite}[1]{\referencename~\onlinecite{#1}}
\DeclareMathOperator{\Tr}{Tr}

\newcommand{\Order}{\mathcal{O}}
\newcommand{\Ns}{\text{N}_\text{s}}
\newcommand{\Nt}{\text{N}_\tau}

\newcommand{\Nf}{\text{N}_\text{f}}
\newcommand{\lattice}[2]{#1^{\vphantom{1}}\times#2^3}
\newcommand{\Lcsc}{L-CSC}
\newcommand{\qcd}{QCD}
\newcommand{\clqcd}{CL\kern-.25em\textsuperscript{2}QCD}

\crefname{figure}{Figure}{Figures}
\crefname{table}{Table}{Tables}
\crefname{equation}{Eq.}{Eqs.}
\crefname{section}{Section}{Sections}
\crefformat{appendix}{the #2Appendix#1#3} % For single Appendix use this to avoid space after label
\newcommand{\mpi}{m_{\pi}}
\newcommand{\PartFunc}{\mathcal Z}
\newcommand{\Action}{\mathcal S}
\newcommand{\SGluon}{\Action_{\text{g}}}
\newcommand{\PL}{Polyakov loop}
\newcommand{\Poly}{L}
\newcommand{\PolyIm}{\Poly_\text{Im}}

\newcommand{\mud}{\hat{m}_{u,d}}
\newcommand{\tauInt}{\tau_{\text{\textsc{int}}}}
\newcommand{\skewness}{B_3}
\newcommand{\kurtosis}{B_4}
\newcommand{\Tc}{T_\text{c}}
\newcommand{\betaC}{\beta_{\text{c}}}
\newcommand{\bnu}{\bar\nu}
\newcommand{\bbetaC}{\bar{\beta}_{\text{c}}}
\newcommand{\MuI}{\hat{\mu}_i}
\newcommand{\MuIpu}{\mu_i}
\newcommand{\RW}{Roberge-Weiss}
\newcommand{\mTric}[1][u,d]{m_{#1}^{\text{tric}}}
\newcommand{\mTricHat}[1][u,d]{\hat{m}_{#1}^{\text{tric}}}
\begin{document}
%================================================================%
\title{%
    Finite size and cut-off effects on the Roberge-Weiss transition in \texorpdfstring{$\Nf=2$}{Nf=2} QCD with Staggered fermions
}
%----------------------------------------------------------------%
\author{Owe Philipsen}
\email{philipsen@itp.uni-frankfurt.de}
\affiliation{
    Institut f\"{u}r Theoretische Physik - Goethe-Universit\"{a}t, Germany \\
    Max-von-Laue-Str.\ 1, 60438 Frankfurt am Main
}
\affiliation{
    John von Neumann Institute for Computing (NIC)
    GSI, Planckstr.\ 1, 64291 Darmstadt, Germany
}
%----------------------------------------------------------------%
\author{Alessandro Sciarra}
\email{sciarra@itp.uni-frankfurt.de}
\affiliation{
    Institut f\"{u}r Theoretische Physik - Goethe-Universit\"{a}t, Germany \\
    Max-von-Laue-Str.\ 1, 60438 Frankfurt am Main
}
%----------------------------------------------------------------%
\date{September 27, 2019}
%================================================================%
\begin{abstract}
    In the absence of a genuine solution to the sign problem, lattice studies at imaginary quark chemical potential are
    an important tool to constrain the QCD phase diagram. We calculate the values of the tricritical quark masses in the 
    Roberge-Weiss plane, $\mu=\imath\pi T/3$, which separate mass regions with chiral and deconfinement phase transitions from the
    intermediate region, for QCD with $\Nf=2$ unimproved staggered quarks on $\Nt=6$ lattices.
    A quantitative measure for the quality of finite size scaling plots  is developed, which significantly reduces the subjective
    judgement required for fitting. We observe that larger aspect ratios are necessary to unambiguously determine
    the order of the transition than at $\mu=0$.
    Comparing with previous results from $\Nt=4$ 
    we find a $\sim\!50$\% reduction in the light tricritical pion mass. The heavy tricritical pion mass stays roughly
    the same, but is too heavy to be resolved on $\Nt=6$ lattices and thus equally afflicted with cut-off effects.
    Further comparison with 
    other discretizations suggests that current cut-off effects on the light critical
    masses are likely to be larger than $\sim\!100$\%, implying a drastic shrinking of the chiral first-order region to possibly zero.
\end{abstract}
%================================================================%
\maketitle
%================================================================%
\section{Introduction}\label{sec:introduction}

The theoretical prediction of the QCD phase diagram as a function of temperature $T$ and baryon chemical potential $\mu_B$
has proved to be a difficult challenge
for several decades. Because of the non-perturbative nature of the strong interactions on
hadronic scales, a first principles approach such as lattice QCD is required. On the other hand, because
of the severe sign problem of lattice QCD at finite $\mu_B$, standard Monte Carlo simulations are limited to addressing
small densities, $\mu_B< 3T$, only~\cite{Philipsen:2010gj,Ratti:2019tvj}. Even at zero baryon density, there remain open questions.
While the thermal transition from a hadron gas to a quark gluon plasma is well established to be an analytic crossover
for physical quark masses~\cite{Aoki:2006we}, the universality class of the transition in the chiral limit of the $u,d$-quarks is still not
settled since it cannot be simulated directly.

\begin{figure}
    \centering
    {\includegraphics[width=0.48\textwidth]{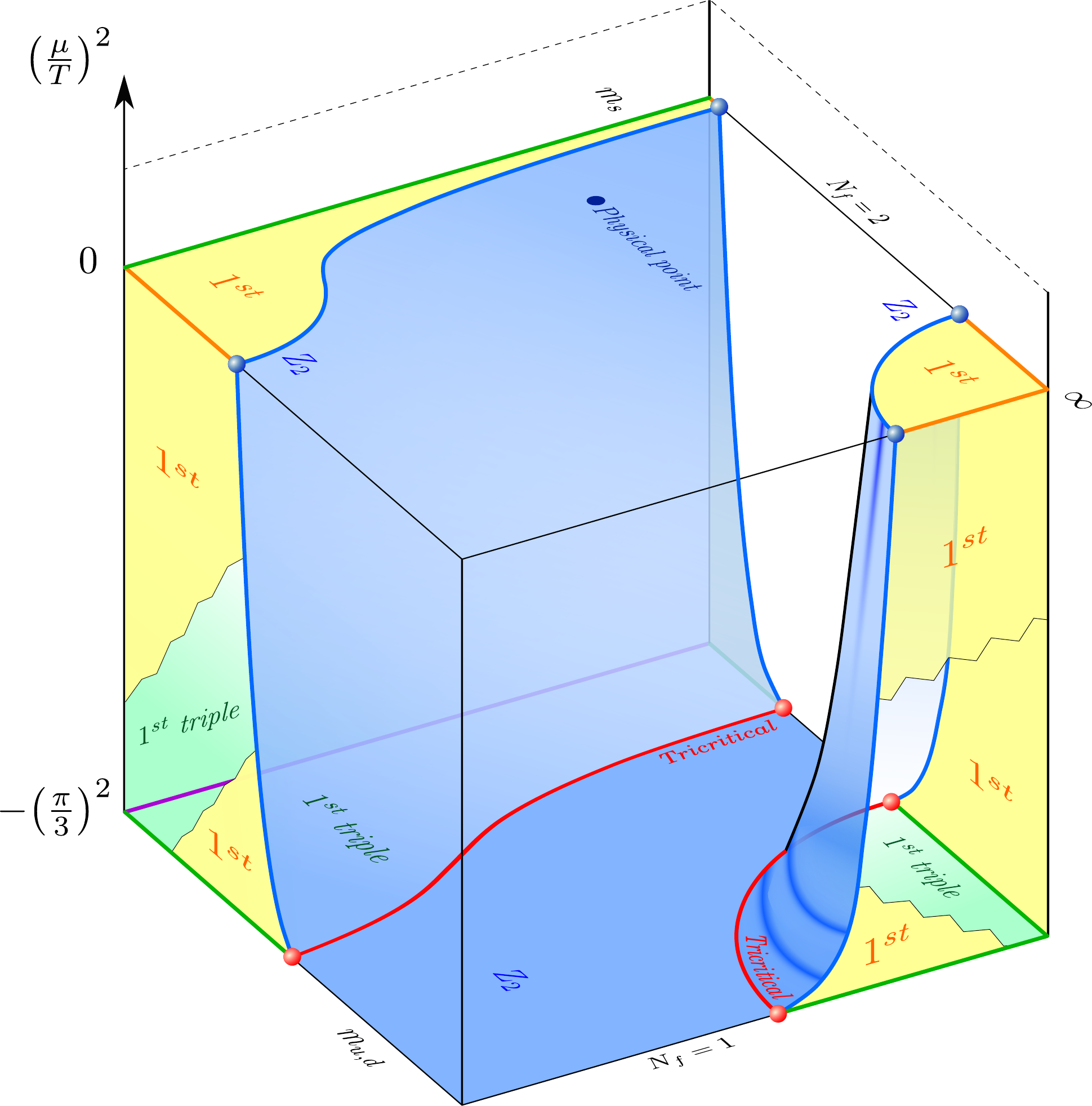}}
    \caption{
      Qualitative sketch of the three-dimensional Columbia plot realized on coarse lattices.
        Whether the chiral, first-order triple region in the \RW\ plane shrinks on finer lattice enough to make an $O(4)$-region appear in the $m_{u,d}=0$ plane remains unclear.
    }
    \label{fig:CP3dI}
\end{figure}

For these reasons, 
it is useful to study the dependence of the thermal transition on QCD parameters like quark masses, numbers of flavors,
imaginary chemical potential, for which there is no sign problem, as well as on the lattice spacing.
The current knowledge of the nature of the QCD thermal
transition as a function of the three light quark masses and imaginary chemical potential, as obtained on coarse lattices
with unimproved actions, is sketched in \cref{fig:CP3dI}. For large and small quark masses, there are regions with
first-order deconfinement and chiral phase transitions, which in the infinite and zero mass limits 
are associated with the breaking and restoration of the center and chiral symmetries, respectively. These are
separated by surfaces of second order transitions  from a large region where the transition is merely an analytic crossover,
to which also QCD with physical parameters belongs~\cite{Aoki:2006we,Bonati:2018fvg}. Note that this qualitative picture is the same for unimproved staggered~\cite{Bonati:2012pe,Cuteri:2017gci}
and unimproved~\cite{Philipsen:2016hkv} as well as improved~\cite{Jin:2017jjp} Wilson discretizations, 
whereas the precise location of the boundary at $\mu=0$ differs significantly
between them, indicating large cut-off effects.
These are also observed for $\Nf=4$ staggered fermions without rooting~\cite{deForcrand:2017cgb}.
By contrast, simulations with improved staggered actions do not see any region of first-order chiral transitions within the available
mass range, neither at zero~\cite{Bazavov:2017xul} nor imaginary chemical potential~\cite{Bonati:2018fvg,Goswami:2018qhc} thus providing 
upper bounds on the critical mass values. 

In the present work we continue earlier studies using the unimproved staggered discretization at imaginary chemical potential on
finer lattices. In particular, referring to \cref{fig:CP3dI}, we investigate how the (red) tricritical points on the $\Nf=2$ line in the 
Roberge-Weiss-plane (bottom plane at $(\mu/T)^2=-(\pi/3)^2$) 
move as the lattice spacing is reduced to $\sim\!2/3$ of its previous values. Together with similar investigations at $\mu=0$, this establishes the
behavior of the critical surfaces when approaching the continuum. Such studies are complementary to ones with improved actions, where
no non-analytic chiral transition is seen, and necessary, if all discretizations are to be understood in the same manner, with expected
agreement in an eventual continuum limit.
As a by-product of our study, we develop a new analysis
of the finite size scaling of cumulants, which significantly reduces the amount of subjective judgement required for fitting.

In order to render the paper self-contained, we briefly summarize the main features of QCD at imaginary chemical potential in \cref{sec:imagmu}. We then proceed to describe our numerical methodology in 
\cref{sec:numericSetup} and our novel analysis method in \cref{sec:collapse}. Our numerical results are given in
\cref{sec:results} before we conclude in \cref{sec:summary}.

%================================================================%
\section{QCD at imaginary chemical potential}\label{sec:imagmu}

\begin{figure}[t]
  \centering
  \includegraphics[scale=0.5]{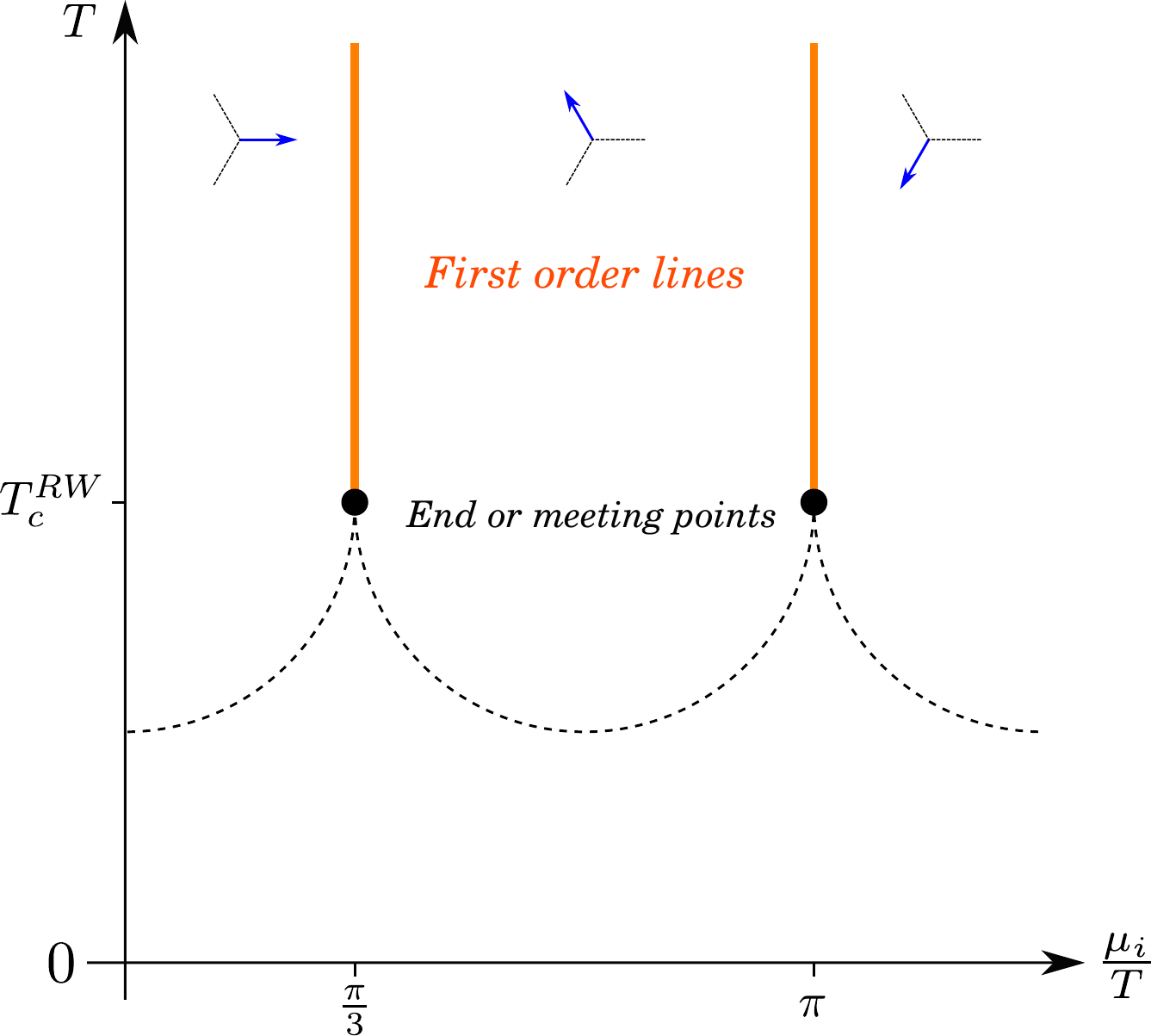}
  \caption{QCD phase diagram in the $T-\MuI$ plane. The dashed line depicts the chiral/deconfinement
           transition whose nature depends on the quark masses. The orange lines represent the Roberge-Weiss
           (RW) transitions. The black dots, where the first-order lines terminate, can be first-order
           triple points, tricritical points or second-order endpoints.}
  \label{fig:RW_T_mu_plane}
\end{figure}

Because of charge conjugation symmetry and its explicit breaking by a non-vanishing baryon density,
the QCD partition function is an even function of quark chemical potential, 
$Z(\mu)=Z(-\mu)$. For purely imaginary chemical potential, $\mu=\imath\MuIpu, \MuIpu\in\mathbb{R}$,
it is furthermore periodic~\cite{Roberge:1986mm},
\begin{equation}
Z(\MuIpu/T)=Z(\MuIpu/T+2\pi k/N_c), \quad k=0,\ldots N_c-1\;,
\end{equation}
and we use $N_c=3$ colors for the QCD gauge group. 
These symmetries imply the phase structure shown in \cref{fig:RW_T_mu_plane}, with three
different $\mathbb{Z}(3)$ center sectors, which are periodically repeated for higher $\MuIpu$. Physical observables,
and in particular the thermodynamic functions, are invariant under a change of sectors, which are 
characterized by different phases of the Polyakov loop
\begin{equation}\label{eq:poly}
L(\mathbf{x})=\frac{1}{3}\mathrm{Tr}\prod_{\tau=1}^{N_\tau-1}U_4(\tau,\mathbf{x})\equiv |L(\mathbf{x})|e^{-i\varphi}\;,
\end{equation}
with $\langle \varphi\rangle =2k\pi/3, k\in\{0,1,2\}$. At high temperatures, there are first-order phase transitions between
the center sectors, whereas at low temperatures they are analytically connected. The dotted line represents the analytic
continuation of the thermal transition, whose order depends on the quark masses. For large and small quark masses,
these lines represent first-order deconfinement and chiral transitions, respectively, whereas for intermediate quark masses they
correspond to an analytical crossover. Consequently, there are three possibilities for the end-point of the Roberge-Weiss transition:
for large and small quark mass it is a first-order triple point, where the thermal first-order transition lines meet that of the center transition.
For intermediate quark masses, the thermal transition is only a crossover
and the center transition ends in a critical end-point in the $3$D Ising universality class. 
At the boundaries between these situations, corresponding to specific quark mass values, the end-point is tricritical 
and corresponds to the red boundary points in the Roberge-Weiss plane of \cref{fig:CP3dI}.
The purpose 
of the present work is to locate these tricritical masses on $N_\tau=6$ lattices with $\Nf=2$ and compare their values with previous determinations
on a coarser $N_\tau=4$ lattice~\cite{Bonati:2010gi}, as well as with those of other discretization schemes.

%================================================================%
\section{Numerical setup}\label{sec:numericSetup}

We consider the QCD partition function of $\Nf=2$ mass-degenerate quarks with a purely imaginary chemical potential. After integration over the fermionic fields it can be written as
\begin{equation}
    \PartFunc(T,\MuI)=\int \mathcal{D}U \;\bigl(\det D[U,\MuI]\bigr)^{1/2} \,e^{-\SGluon[U]}\;,
\end{equation}
where $\SGluon$ is the gauge part of the action and $D$ is the fermion matrix.
For our investigation we used the standard Wilson gauge action and the standard staggered discretization of dynamical fermions.
Denoting the lattice gauge coupling by $\beta=6/g^2$, with the continuum gauge coupling $g$, and an elementary plaquette by $P$, 
we have
\begin{equation}
    \SGluon[U]= \beta \sum_{P}\Bigl\{ 1 - \Re \bigl[\Tr_C P\bigr]\Bigr\} \;.
\end{equation}
The fermion matrix reads
\begin{align}
    D_{i,j} &= \mud\,\delta_{i,j} \;+ \notag\\
            &+ \frac{1}{2}\sum_{\nu=1}^4 \eta_{i,\nu}
               \Bigl(\tilde{U}_{i,\nu}\,\delta_{i,j-\hat\nu} - \tilde{U}^{\dagger}_{i-\hat\nu,\nu}\,\delta_{i,j+\hat{\nu}}\Bigr)\;,
\end{align}
where $\mud=am_{u,d}$ is the quark bare mass in lattice units, $a$ is the lattice spacing, $i,j$ refer to lattice sites, $\eta_{i,\nu}$ are the staggered phases, $\hat\nu$ is a unit vector on the lattice and $\tilde{U}_{i,\nu}$ are the gauge links, which include the purely imaginary chemical potential $\MuI=a\MuIpu$ in the temporal direction,
\begin{equation}
    \tilde{U}_{i,\nu}=\left\{
    \begin{aligned}
        U_{i,\nu} &\qquad\nu\in\{1,2,3\} \\
        e^{i\MuI}U_{i,\nu} &\qquad\nu=4
    \end{aligned}
    \right.\quad.
\end{equation}
The temperature is specified by the inverse euclidean time extent of the lattice,
\begin{equation}
    T=\frac{1}{a(\beta)\:\Nt}\;.
\end{equation}

\begin{table}[t]
  \centering
  \[
  \begin{array}{*{5}{c}}
    \toprule[0.3mm]
    & \text{Crossover} & 1^{st} \text{ triple} & \text{Tricritical} & 3\text{D Ising} \\
    \midrule[0.1mm]
    \kurtosis    & 3 & 1.5 & 2   & 1.604     \\
    \nu          & - & 1/3 & 1/2 & 0.6301(4) \\
    \gamma       & - & 1   & 1   & 1.2372(5) \\
    \bottomrule[0.3mm]
  \end{array}
  \]
  \caption{Critical values of $\nu$, $\gamma$ and $\kurtosis\equiv\kurtosis(X,\ldots)$
           for some universality classes~\cite{Pelissetto:2000ek}.}
  \label{tab:kurtosisValues}
\end{table}

\begin{figure*}[t]
    \centering
    \subfigure[$m=0.01$]{\label{fig:kurtosisCrossingYES}\includegraphics[width=0.48\textwidth]{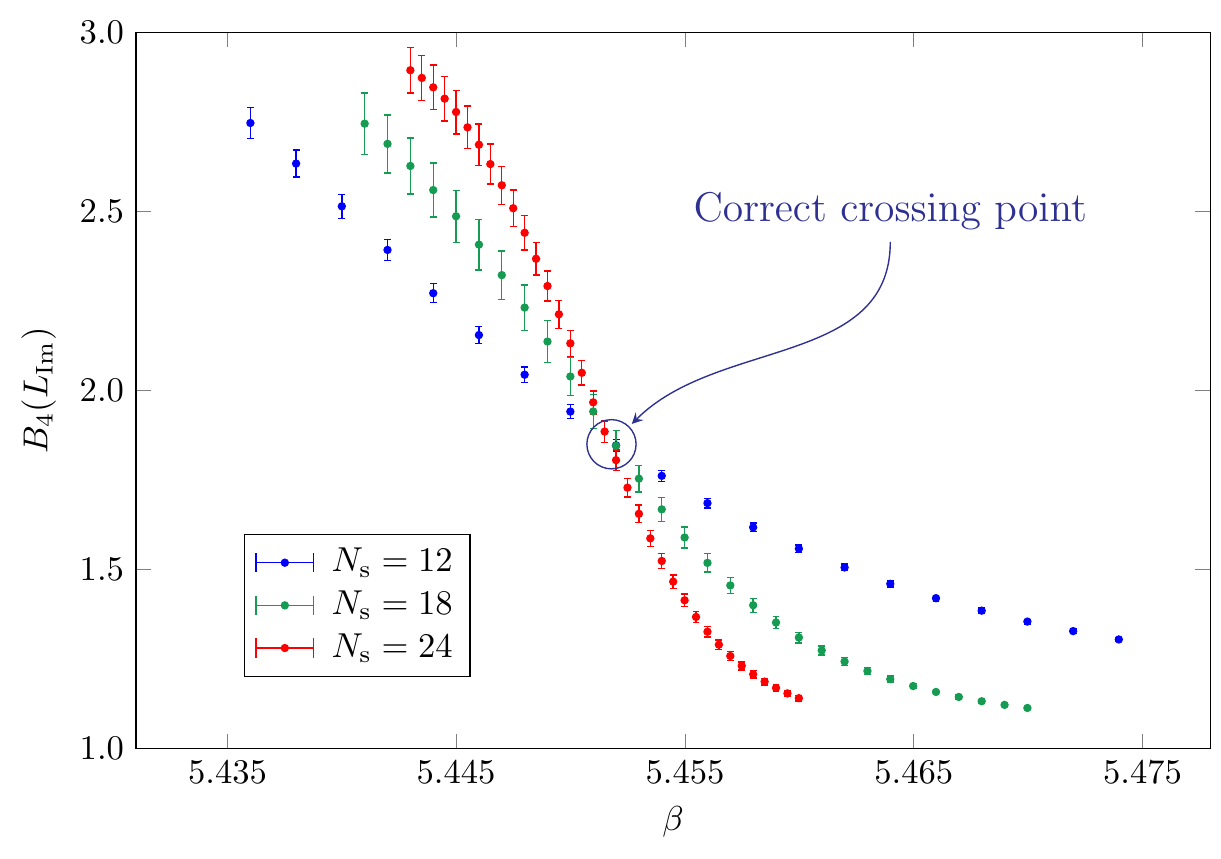}} \quad
    \subfigure[$m=0.45$]{\label{fig:kurtosisCrossingNO}\includegraphics[width=0.48\textwidth]{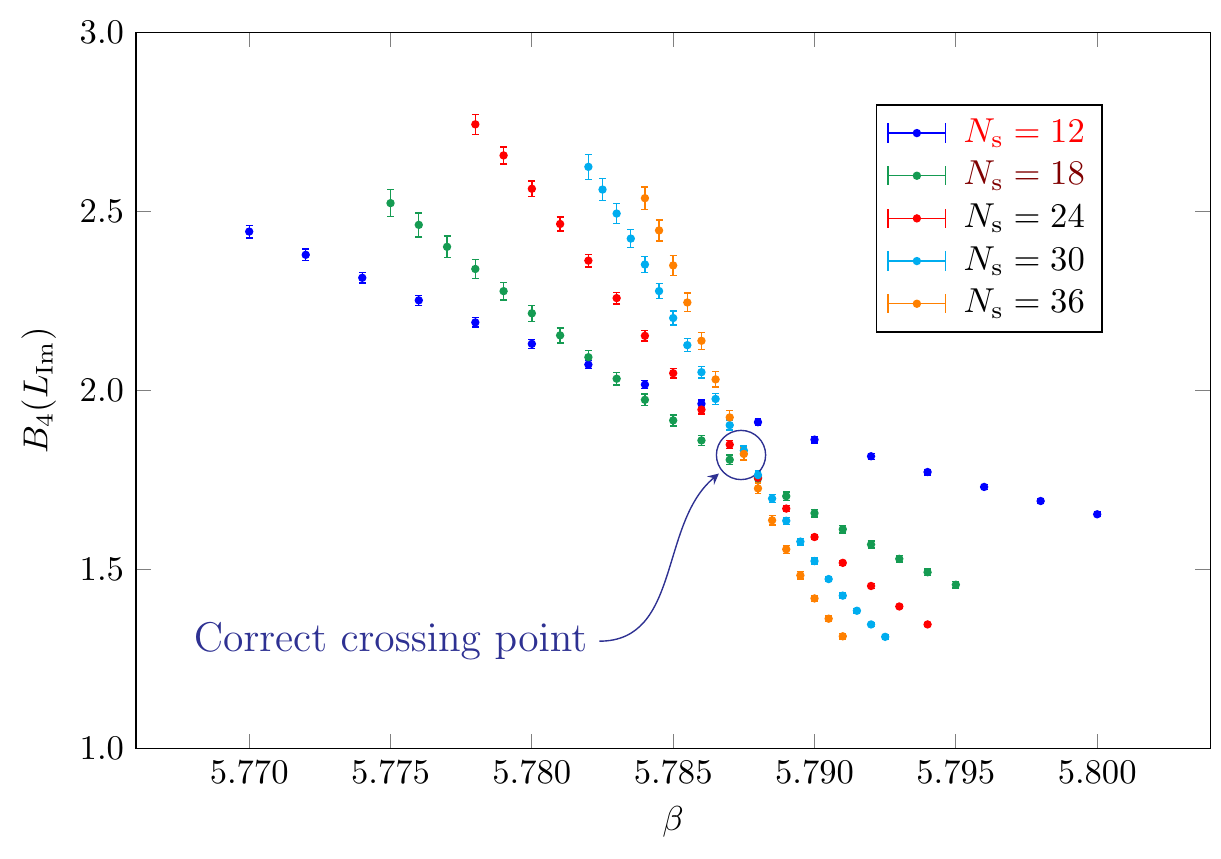}}
    \caption{Kurtosis of the imaginary part of the \PL{} as function of $\beta$ at two different values of the quark mass.
             The plot at $m=0.45$ is a typical example of what can happen when finite size effects are too large.
             Clearly, the data at $\Ns=12$ have not been included in the finite size scaling analysis.}
    \label{fig:kurtosisCrossing}
\end{figure*}

In order to locate a phase transition and study its nature, we calculate standardized cumulants
\begin{equation}\label{eq:standardizedCumulants}
    B_n(X,\beta,\mud,\MuI)
         \equiv\frac{\bigl\langle(X-\langle X\rangle)^n\bigr\rangle}
                    {\bigl\langle(X-\langle X\rangle)^2\bigr\rangle^{\frac{n}{2}} \vphantom{\Bigl[}}\;,
\end{equation}
constructed from an (exact or approximate) order parameter $X$. 
In particular, a non-trivial zero of the skewness of the $X$-distribution,
\begin{equation}\label{eq:zeroSkew}
    \skewness(\betaC)=0\;,
\end{equation}
determines at which value of $\beta=\betaC$ a thermal transition takes place, while the value of the kurtosis $\kurtosis$ 
in the thermodynamic limit, evaluated at the critical coupling, will determine the order of the phase transition 
(refer to~\cref{tab:kurtosisValues} for common kurtosis values).
Note that the so-called Binder cumulant~\cite{Binder:1981sa},
\begin{equation}
    \mathcal{U}_4(X)\equiv 1-\frac{\bigl\langle(X-\langle X\rangle)^4\bigr\rangle}
                                 {3\bigl\langle(X-\langle X\rangle)^2\bigr\rangle^2 \vphantom{\Bigl[}}
                     =1-\frac{1}{3}\,B_4\;,
\end{equation}
is trivially related to the kurtosis (of the same observable) and contains the same information.

We fix $\MuIpu/T=\pi$ since in this case the imaginary part of the \PL\ is an exact order parameter,
\begin{equation}
X=\PolyIm\equiv \frac{1}{\Ns^3}\sum_\mathbf{x}\mathrm{Im}(L(\mathbf{x}))\;.
\end{equation}
Referring to \cref{eq:poly}, it distinguishes between the low $T$ disordered phase and the high $T$ ordered phase with two-state coexistence,
\begin{equation}
    \left\{
    \begin{aligned}
        \vphantom{x^\frac{1}{2}}\langle\varphi\rangle     =0 \;\Rightarrow\; \langle \PolyIm\rangle     =0  \quad& \text{low }T\\
        \vphantom{x_\frac{1}{2}}\langle\varphi\rangle \neq 0 \;\Rightarrow\; \langle \PolyIm\rangle \neq 0  \quad& \text{high }T\\
    \end{aligned}
    \right.\quad,
\end{equation}
with the advantage of knowing its mean value exactly.

On a $\Nt=6$ lattice, we thus set $\MuI=\pi/6$. Since this is the boundary between two Roberge-Weiss sectors for all temperatures, 
$\skewness(\PolyIm)=0$ for any value of $\beta$ and we cannot use \cref{eq:zeroSkew} to locate the \RW\ end-point.
However, the kurtosis $B_4$ is expected to vary from values close to $3$ (crossover) at low $T$ to values close to $1$ (first order) at high $T$.
Although it becomes a non-analytic step function in the $V\to\infty$ limit, it is a smooth function on finite volumes,  with the curves
for different volumes crossing at a universal value for  $B_4$ 
at the critical point $\beta=\betaC$, provided that the spatial lattice extent is large enough.
This crossing provides the location of the \RW\ end-point.
In the neighborhood of the critical point $\betaC$, the kurtosis shows a well-defined finite size scaling behavior
as a function of the scaling variable
\begin{equation}\label{eq:scalingVariable}
    x\equiv(\beta-\betaC)\Ns^{1/\nu}\;.
\end{equation}
Its Taylor expansion around the critical point $x=0$ is 
\begin{equation}\label{eq:kurtosisExpansion}
    \kurtosis(\beta, x) = \kurtosis(\betaC,\infty) + a_1\,x + a_2\,x^2 + \Order(x^3) \;.
\end{equation}
Sufficiently close to the thermodynamic limit, the coefficient $\kurtosis(\betaC,\infty)$ and the critical exponent $\nu$ take their universal values depending on the type of transition.

In order to locate the two tricritical points in the \RW\ plane, we performed simulations at different values of $\mud$ and  different values of 
$\beta$ around the critical temperature.
Evaluating the kurtosis in the critical region and fitting it to \cref{eq:kurtosisExpansion}, considering the linear term only, gives 
$\kurtosis(\betaC,\infty)$, $a_1$, $\betaC$ and $\nu$ for every value of $\mud$.
The change of $\nu$ as a function of $\mud$ then permits to locate the light and heavy $\mTricHat$ values.

Although our main quantitative analysis is based on the kurtosis of the order parameter, we also calculated the susceptibility of $|L|$,
\begin{equation}
    \chi(|L|)\equiv \Ns^3 \bigl\langle(|L|-\langle |L|\rangle)^2\bigr\rangle   \;.
\end{equation}
which is expected to scale around $\betaC$ according to
\begin{equation}
    \chi=\Ns^{\gamma/\nu} f(t\,\Ns^{1/\nu})   \;.
\end{equation}
Here $t\equiv(T-\Tc)/\Tc$ is the reduced temperature and $f$ is a universal scaling function.
Comparing the collapse plots obtained by fixing the critical exponents $\gamma$ and $\nu$ to the first-order or second-order values, 
and by plotting $\chi/\Ns^{\gamma/\nu}$ evaluated on different lattice sizes against $t\Ns^{1/\nu}$ also provides information
about the nature of the thermal transition and serves as a cross-check of the kurtosis analysis.
A similar cross-check using the chiral condensate $\langle\bar\psi\psi\rangle$ was occasionally performed in the small-mass region, 
leading to consistent conclusions.

We investigated $19$ values of $\mud$ in the intervals $[0.004, 0.011]$ and $[0.15,0.85]$.
For each value of $\mud$, three to five spacial lattice sizes have been used, keeping $\Nt=6$ and $\MuI=\pi/6$ fixed.
This corresponds to aspect ratios $\Ns/\Nt\in[2,7]$.
Larger spatial volumes than initially chosen were added whenever the kurtosis of the order parameter on different volumes was 
not crossing at the same point (an example is reported in \cref{fig:kurtosisCrossing}).
For every lattice size, between three and seven values of $\beta$ around the critical temperature have been simulated.
In between those, the observables have been evaluated at additional $\beta$-values using the Ferrenberg-Swendsen multiple histogram method (also known as reweighting technique)~\cite{Ferrenberg:1989ui} to increase resolution (see the discussion at the end of this and in the next section).

Configurations were generated by a standard RHMC algorithm~\cite{Kennedy:1998cu}, 
producing four different Monte Carlo chains per $\beta$ with unit-length trajectories.
Where advantageous ($\mud<0.007$), the multiple-pseudofermions technique~\cite{Clark:2006fx} has been used.
The algorithm acceptance has been tuned to be not lower than $80\%$.
At least $5k$ trajectories were always discarded as thermalization and afterwards observables of interests (i.e. the plaquette, the \PL\ and the chiral condensate for small masses) have been computed for every trajectory.
We increased statistics until the standard deviation of the kurtosis $\kurtosis(X)$ decreased below $\sim0.2$, and $\kurtosis(X)$ was the same on all the four chains at the same $\beta$ within two (three) standard deviations in the large (small) mass region.
For this reason the collected statistics per $\beta$ is not uniform, detailed information is given in \cref{app:sim}.
In order to satisfy these strict requirements, millions of trajectories per volume and almost half a billion in total have been produced.
This large statistics is necessary due to the large autocorrelation times shown by $\PolyIm$, especially when entering the first-order regions (cf.~\cref{tab:statistics} in \cref{app:sim}).
We always insisted on having at least $100$ independent events per $\beta$ in the analysis.

In order to determine the lattice spacing and the pion mass, also zero-temperature simulations have been performed.
We produced $800$ independent configurations on $\lattice{32}{16}$ lattices for each value of $\mud$.
Using the publicly available code described in \refcite{Borsanyi:2012zs}, the scale was set by the Wilson flow parameter $w_0$.
Pion masses, instead, were measured with standard spectroscopy techniques~\cite{Golterman:1985dz,Ishizuka:1993mt}.

All our numerical simulations (except those for scale setting purposes) have been run using the publicly available~\cite{CL2QCD} \texttt{OpenCL} based code \clqcd~\cite{Philipsen:2014mra}, which is optimized for GPUs.
The \Lcsc~\cite{L-CSC} supercomputer at GSI in Darmstadt has been used, and the thousands of jobs needed in the study have been efficiently handled using the simulation monitoring package \texttt{BaHaMAS}~\cite{Sciarra:2017gmt}.

Our quite intricate fitting procedure used to extract the critical exponent $\nu$ is completely analogous to the one previously 
described in Appendix B of~\refcite{Cuteri:2015qkq}.
For each value of the quark mass, nearly all possible fits of the data to the linear part of \cref{eq:kurtosisExpansion} are performed, and a filtering procedure is applied afterwards in order to pick the best fits.
This is needed because the range in which $\kurtosis(\PolyIm)$ can be considered linear for each $\Ns$ value is not known a priori.
However, here we differ in a technical detail from the previous study in~\refcite{Cuteri:2015qkq}.
As already mentioned, reweighting was used not only to smoothen the signal, but also to supply additional $\beta$-points for the fit.
This approach allows to reduce the number of required simulations, provided there are
clear criteria according to which such points are added.
Reweighted points introduce a correlation with the others, and too  many of them would render the fits unreliable. 
Hence, we added more reweighted points between simulated points only if with lower resolution it was not possible to obtain a good fit.
Moreover, another important aspect should be considered in choosing the reweighting resolution.
The $\beta$-region where the kurtosis is linear shrinks on larger volumes.
Thus, choosing the same resolution in $\beta$ on different $\Ns$ would imply to include fewer points from larger volumes in the fit and, 
consequently, to enhance finite size effects.
Therefore, we increased the reweighting resolution in $\beta$ on larger $\Ns$, making the information coming from the smallest volume systematically less important, see \cref{tab:fits} for detailed overview.
This aspect is even more explicit looking at how many points per $\Ns$ have been included in the fit.

%================================================================%
\begin{figure}[t]
    \centering
    \subfigure[First-order critical exponents]{\includegraphics[width=0.48\textwidth]{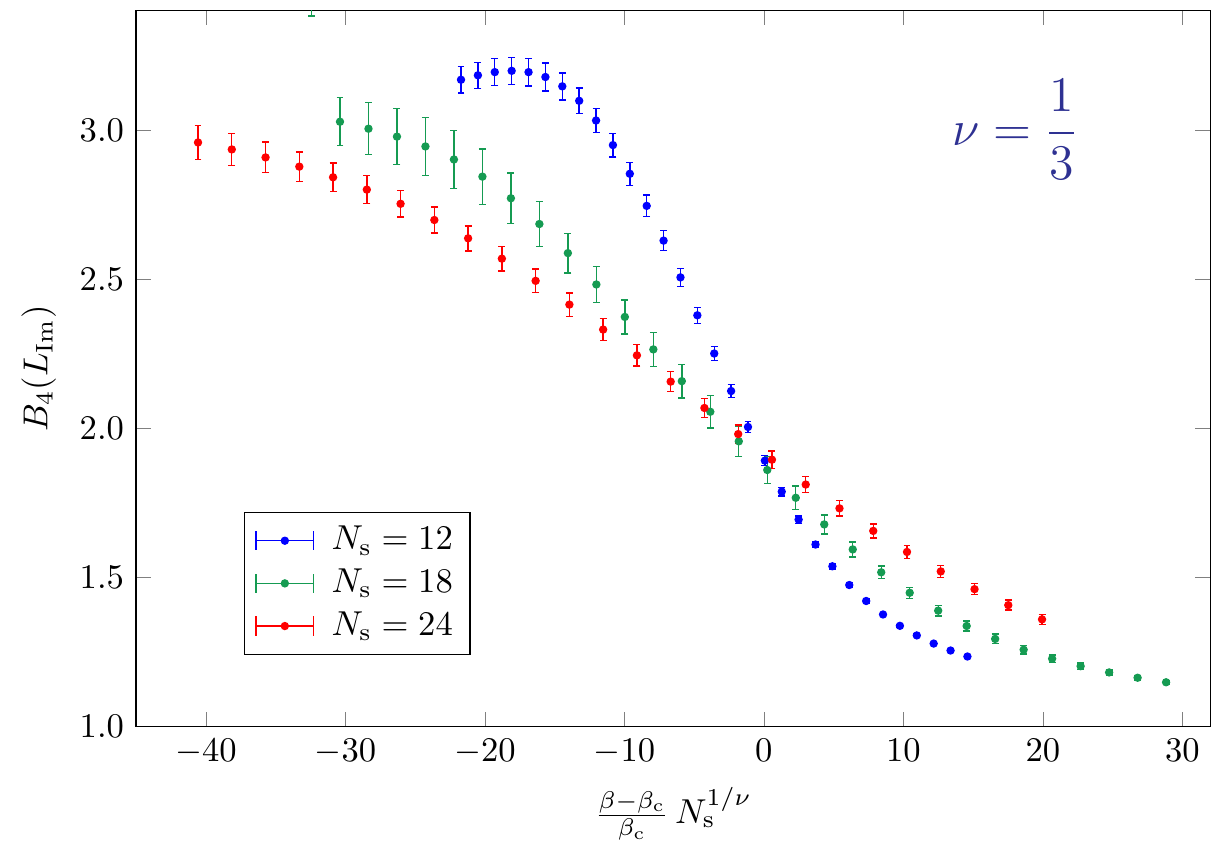}} \\[5mm]
    \subfigure[Second-order critical exponents]{\includegraphics[width=0.48\textwidth]{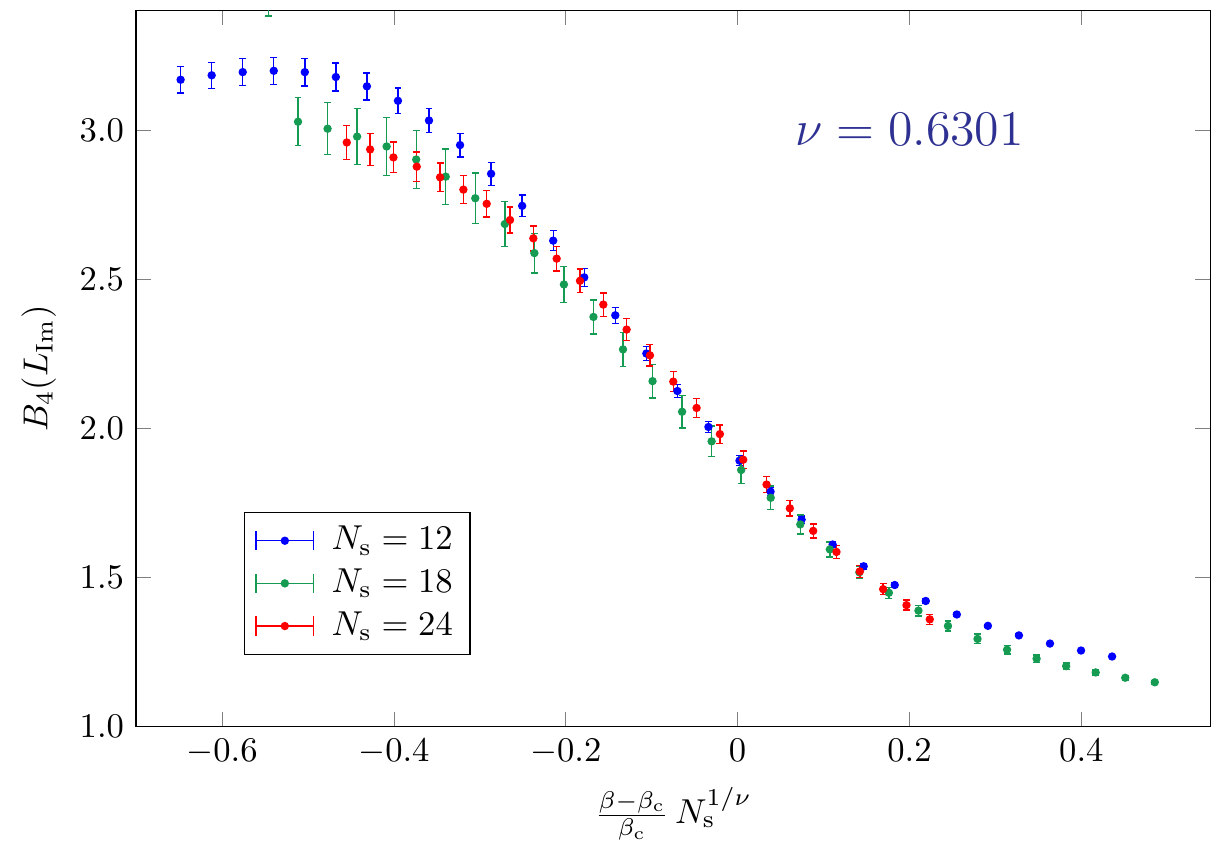}}
    \caption{Collapse plot of the kurtosis of the imaginary part of the \PL{} for $\mud=0.25$ using known values of the critical exponent $\nu$.
             Since this mass is far from the tricritical points, a by-eye judgement is enough to rule out the first order as possible type of phase transition.}
    \label{fig:kurtosisCollapse}
\end{figure}

\section{An alternative to fitting: quantitative collapse plots}\label{sec:collapse}

Any fitting procedure, however careful, relies on a few subjective decisions like the 
number of reweighted points and the filtering parameters to judge a good fit.
We now propose an alternative procedure to independently
determine the critical exponent from scaling/collapse plots, which can then be compared with the results of the fitting procedure. 

\begin{figure*}[t]
    \centering
    \includegraphics[viewport=0 0 340 180, clip, width=\columnwidth]{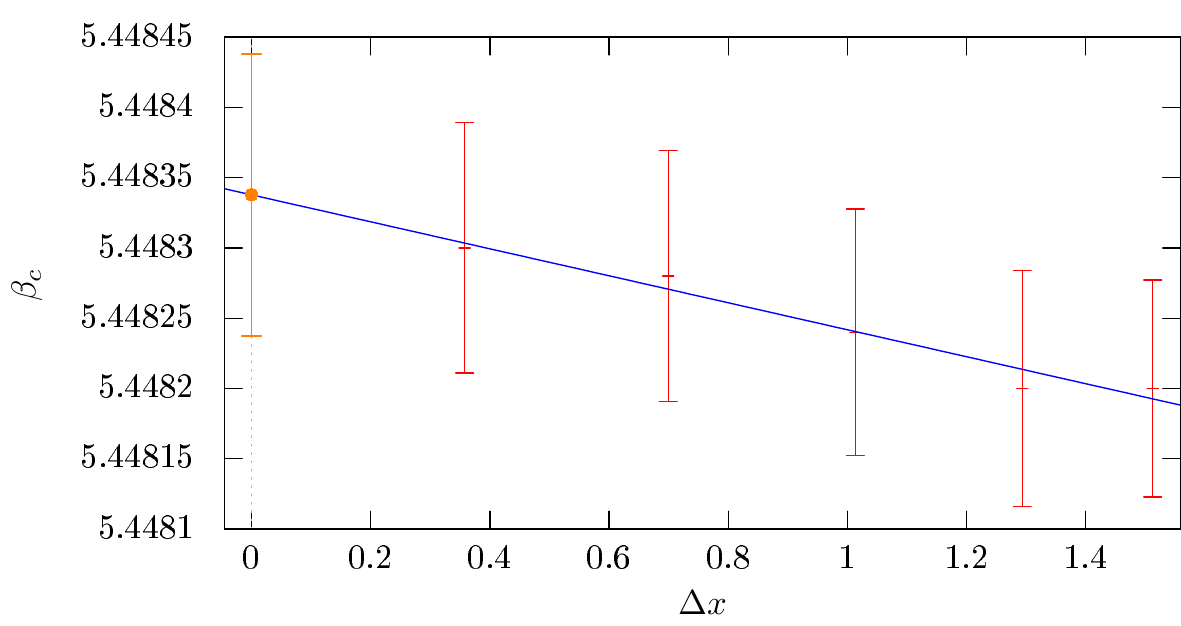}
    \hfill\includegraphics[viewport=0 0 340 180, clip, width=\columnwidth]{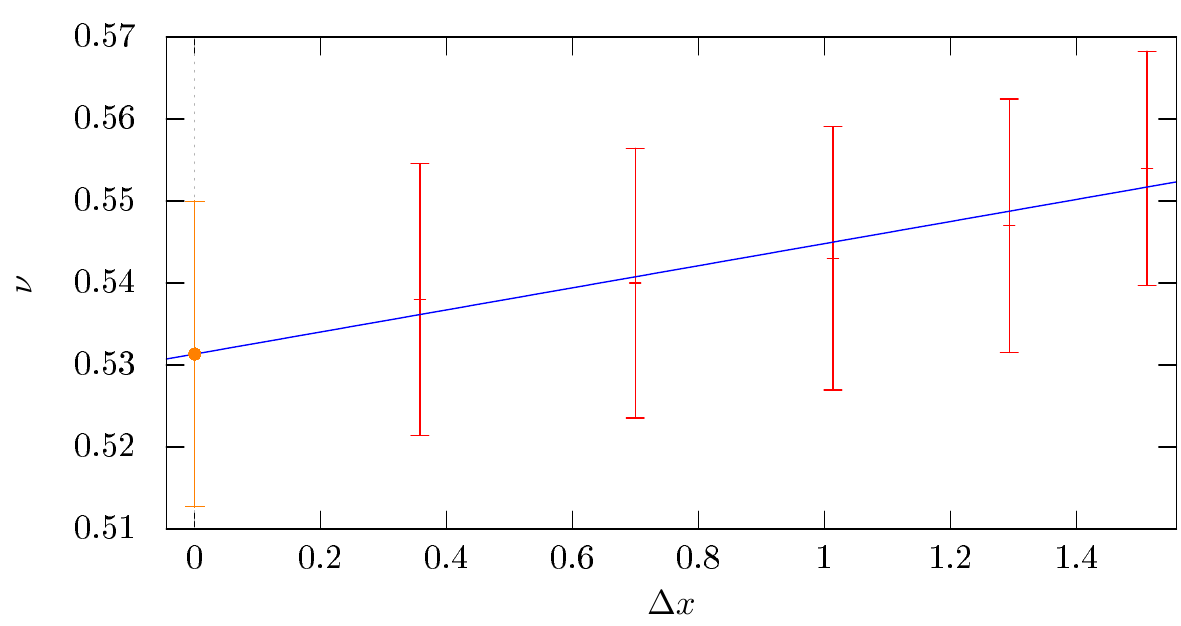}
    \caption{Example of linear extrapolation $\Delta x\to0$ in the quantitative collapse plot analysis for $\mud=0.009$ for $\betaC$ and $\nu$.}
    \label{fig:extrapolationCollapse}
\end{figure*}

A collapse plot is obtained if an observable, which displays universal finite size scaling, is plotted
as a function of its scaling variables, such that the curves for different volumes fall on top of each other provided 
the volumes are sufficiently large to represent the thermodynamic limit.
There are several common observables used for this purpose, 
here we focus on $\kurtosis(\PolyIm)$ as function of the scaling variable $x$ defined in \cref{eq:scalingVariable}.
An example is shown in \cref{fig:extrapolationCollapse}.

Whenever the lattice volume is not large enough, scaling is violated and no good collapse is obtained, even when the known critical values (listed in \cref{tab:kurtosisValues}) are used.
Finite size corrections are responsible for that and, in principle, a better collapse can be obtained using different (non-universal) values of the exponents.
The quality of the collapse is usually judged by eye, 
which is mostly sufficient to distinguish between a first and a second order phase transition with known exponents (like in \cref{fig:kurtosisCollapse}).
However,  a more rigorous method is clearly needed in a situation where the scaling exponents change from 
first to second-order. On any finite volume, this will lead to intermediate values of the exponents, which should
be determined unambiguously together with an associated error. 
For this purpose we now construct a quantitative measure of the collapse of our data.

Considering how we judge a collapse plot by eye, the measure of the quality has to be related to the distance between different points 
at the same value of the universal scaling variable.
Inspired by the method used by Barkema and Newman~\cite{Barkema:1996mc} for the thermal random-field Ising model (see also a similar analysis in Appendix A of~\refcite{Houdayer_2004}), we associate a quantitative quality to the collapse of $\kurtosis(\PolyIm)$ by estimating the average variance of
the data as
\begin{align}\label{eq:qualityCollapse}
    Q(\bbetaC,\bnu)\equiv\frac{1}{\Delta x}
    \int_{x_{\text{min}}}^{x_{\text{max}}}\Biggl\{
    &N_V\sum_{i=1}^{N_V}\Bigl[\kurtosis\bigl(x(\bar{\beta}_c,\bnu,V_i)\bigr)\Bigr]^2 \\
    &- \Biggl[\sum_{i=1}^{N_V} \kurtosis\bigl(x(\bar{\beta}_c,\bnu,V_i)\bigr)\Biggr]^2 \Biggr\} \text{d}{x}\;.\nonumber
\end{align}
Here, $\Delta x\equiv x_{\text{max}}-x_{\text{min}}$ is the considered range in the scaling variable, $V\equiv\Ns$, $N_V$ is the number of considered lattice volumes, while $\bbetaC$ and $\bnu$ are fixed values for the critical temperature and for the critical exponent $\nu$, respectively.
A normalisation factor $N_V^{-2}$ was neglected in front of the expression, since it is irrelevant for the estimate of the critical exponent. 
It is now possible to obtain an estimate for $\betaC$ and $\nu$ by minimizing $Q$ as function of these two variables.
Nevertheless, this is a non-trivial task and there are some problems to be addressed.

The integration in \cref{eq:qualityCollapse} must be done numerically, since the exact functional form of $\kurtosis(x)$ is unknown.
This, in principle, would not be a problem, if only we had the kurtosis at the same $x$ on different volumes.
However, in lattice simulations the kurtosis $\kurtosis$ is measured at fixed values of $\beta$, and the mapping~\eqref{eq:scalingVariable} between $\beta$ and $x$ depends on the unknown parameters $\betaC$ and $\nu$.
Therefore, it is not possible to have simulated data uniformly spaced in $x$ for any $\betaC$ and $\nu$.
On the other hand, the measured data can be interpolated in $\beta$ using the multiple histogram method.
Hence, it is possible to reweight the kurtosis in $\beta$ in such a way that its values at the same $x$ are available for all the volumes.
After this step the calculation of $Q$ is trivial.
In practice, this implies an interpolation for each pair $(\bbetaC,\bnu)$ at which $Q$ has to be evaluated, which is too costly if a precise determination of the final value of the critical exponent is desired.
A cheaper alternative is to use the reweighting technique to obtain the kurtosis as an approximately continuous function of $\beta$, i.e., to add a large number of points between two simulated temperatures.
The numerical integration to obtain $Q$ can then be performed with negligible additional error.
However, due to the particular form of the map $x(\beta)$, sometimes, especially for small values of $\nu$, the number of interpolated points needed to have a sufficiently precise numerical integration can become very large and, therefore, the reweighting very costly.
A smarter approach is then required.

As can be seen in \cref{fig:kurtosisCrossing}, the kurtosis of the imaginary part of the \PL{} is a quite regular function of $\beta$, in the sense that no sudden variations are present.
This means that a numerical interpolation of the data which does not take into account the physics~--~as the multiple histogram method does~--~will probably still find the correct value of the kurtosis.
Clearly, this is true under the assumption that the resolution of the data to be interpolated is high enough.
For example, the simulated data are usually too distant in temperature to be correctly interpolated without the reweighting technique.
But after having applied the multiple histogram method to the data, a second interpolation can be done very cheaply.
In practice, we used the software \texttt{Mathematica} to obtain an interpolated function out of a set of points and perform numeric operations on it.
The advantage of having a kurtosis as a \emph{function} makes the calculation of $Q$ straightforward.
Furthermore, it is then possible to automatically minimize $Q(\betaC,\nu)$ as function of two variables.

Next, we need to estimate the statistical error on $\betaC$ and on $\nu$,
which has to contain the error on the reweighted points and the statistical error of our simulations.
An error on reweighted points is often obtained using the bootstrap method.
This means that, in the reweighting procedure, $N_{\text{boot}}$ sets of reweighted kurtosis values are calculated, and the bootstrap errors are extracted from them.
Now, instead of using these sets to compute errors on the kurtosis, they can be used to minimize $Q$, obtaining $N_{\text{boot}}$ different estimates of $\betaC$ and of $\nu$, which will give the desired final error.
Since, typically, the number of bootstrap samples is of the order of some hundreds, it is clear that the minimization of $Q$ should not take too much time\footnote{In \texttt{Mathematica}, for example, it is possible to use the \texttt{NargMin} function, but a user implemented minimization based on a scan in $\betaC$ and in $\nu$ is more efficient, though less precise.}.

Finally, let us discuss how $x_{\text{min}}$ and $x_{\text{max}}$ should be chosen.
Clearly, no extrapolation outside the simulated interval in $\beta$ should be done.
Thus, the largest $\Delta x$ is the interval in $x$ around $0$ where data from \emph{all} volumes are available.
Since $x_{\text{c}}=0$, we have $x_{\text{min}}<0$ and $x_{\text{max}}>0$ and,
in order to have a symmetric Taylor expansion interval~\cite{Cuteri:2015qkq}, we chose
\begin{equation}
    \lvert x_{\text{min}} \rvert=\lvert x_{\text{max}} \rvert\;.
\end{equation}
Using too large an interval of integration is, in general, not correct, since it assumes data collapse possibly outside the critical region.
On the other hand, the width of the scaling region is not known a priori.
A clever solution to this problem, successfully applied in~\cite{Barkema:1996mc}, consists of repeated analyses 
for successively decreasing $\Delta x$ followed by an extrapolation of the resulting parameters to $\Delta x\to0$.
An example is reported in \cref{fig:extrapolationCollapse}.

%================================================================%
\begin{figure*}[t]
    \centering
    \includegraphics[width=0.95\textwidth, clip, trim=3mm 0 0 0]{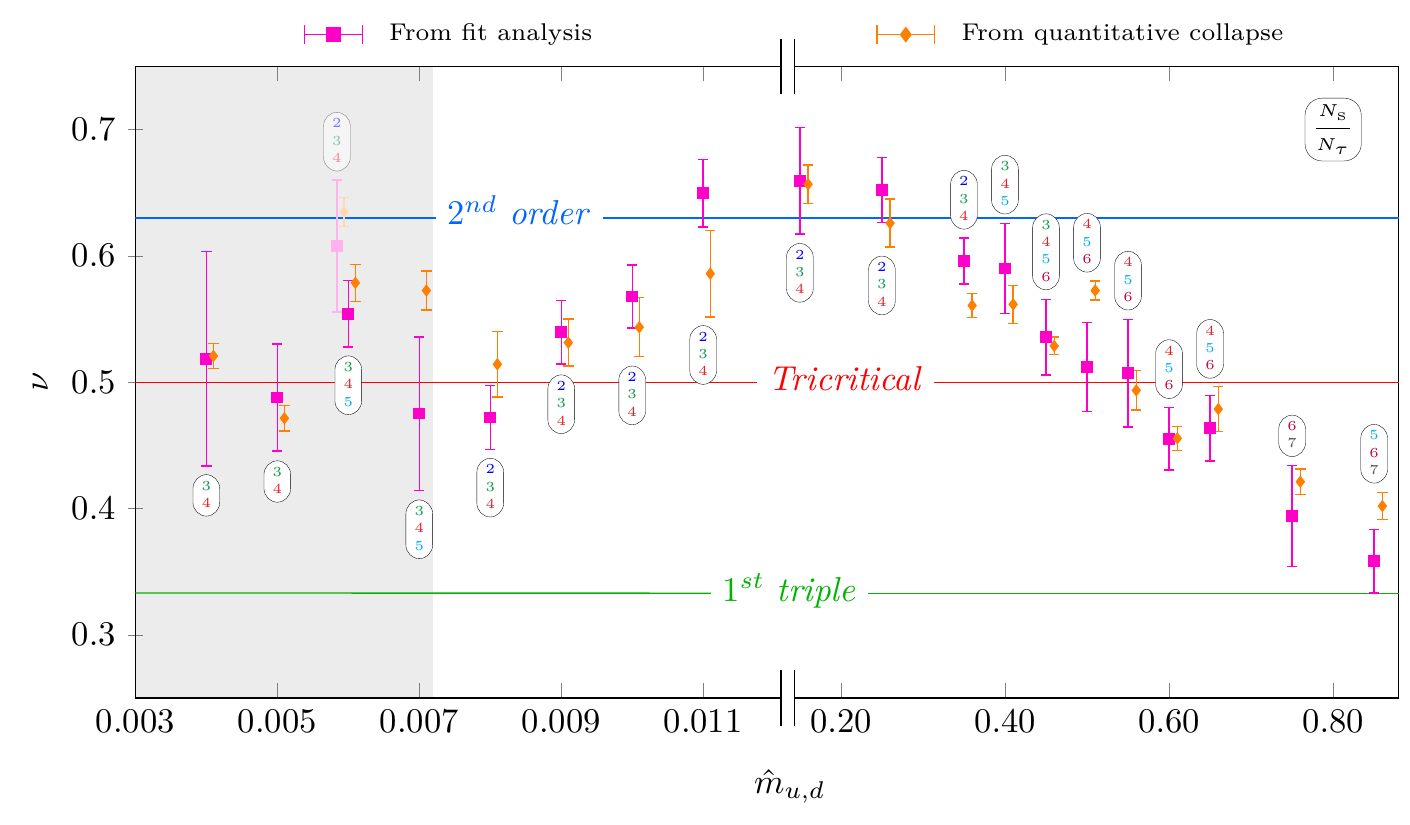}
    \caption{Critical exponent $\nu$ as function of the bare quark mass $\mud$ obtained with two different methods.
             The quantitative collapse plot points have been slightly horizontally shifted to avoid superposition (refer to \cref{tab:fits,tab:collapse} for the the data).
             At $\mud=0.006$ the outcome of the data analysis without the largest spatial value (shaded points) has been included to guide the discussion in the text.
             For each mass, a badge containing the aspect ratios $\Ns/\Nt$ used in the analysis has been drawn.
             The horizontal colored lines are the critical values of $\nu$ for some universality classes.
             The mass axis has been broken and two different scales have been used in order to improve readability.
             Points in the grey-background region have to be taken with a pinch of salt, since it is reasonable to believe that finite size effects are for them dominant.}
    \label{fig:NuMass}
\end{figure*}

\section{Numerical results and discussion}\label{sec:results}

As mentioned in \cref{sec:numericSetup}, our strategy to locate the tricritcal points is to measure the critical exponent $\nu$ for different quark masses and see where it changes from its first-order
value $1/3$ for small and large masses to the $3$D Ising value \si{0.6301(4)} for intermediate masses.
The changes approach a step function in the thermodynamic limit but remain smooth as far as finite lattice volumes 
are used to extract $\nu$.
The critical exponent $\nu$ is preferable over $\kurtosis(\betaC, \infty)$, since it is known to suffer less from finite volume corrections~\cite{deForcrand:2010he,Philipsen:2014rpa,Cuteri:2015qkq}.
The main result of our investigation is reported in \cref{fig:NuMass} (more detailed information about the displayed data is available in \cref{tab:collapse,tab:fits} in \cref{app:sim}).
For each value of the quark mass, the critical exponent $\nu$ is extracted, both with the fit analysis used already in \refcite{Cuteri:2015qkq} and with the new collapse strategy introduced in \cref{sec:collapse}.
The agreement between the two methods to extract the critical exponent $\nu$ is evident in \cref{fig:NuMass}.
The quantitative collapse analysis has systematically smaller errors on $\nu$, though.
This, together with the fact that no arbitrary decision in the analysis may affect the outcome, should make this method preferable.

In the large mass region, the signal is quite smooth and $\nu$ changes monotonically from the second-order to the first-order value.
However, approaching and entering the first-order region, the minimal aspect ratio $\Ns/\Nt$ needed to extract $\nu$ increases significantly from $2$ to $6$ compared to studies at $\mu=0$.
This has been remarked already in previous studies~\cite{Bonati:2010gi,Cuteri:2015qkq}.
Presumably this is due to the fact that in the Roberge-Weiss plane we are dealing with the more complex 
three-state coexistence and its coalescence in a tricritical point. 
 
The same behavior is expected also in the small-mass first-order region, 
where simulations with an aspect ratio larger than $4$ are too costly.
Therefore, in \cref{fig:NuMass}, the mass region $\mud\le 0.007$ has been marked with a gray background to stress that larger volumes 
are required to polish the result.
However, we have reasons to believe that the tricritical point is already located.
It is always possible to compare the critical exponent $\nu$ extracted using only part of the available volumes, while leaving
the smallest or largest out of the analysis.
In this way finite size effects are made visible by checking whether $\nu$ drifts towards first-order or second-order values upon
inclusion of larger volumes.
This is shown for $\mud=0.006$ in \cref{fig:NuMass} where a clear decrease in $\nu$ is visible adding $\Ns=30$ and removing $\Ns=12$ 
in the analysis.
Another aspect that made us confident to be entering the first-order region for $\mud\lesssim0.007$ is the typical ``Binder bump'' behavior discussed in detail in~\refcite{Cuteri:2015qkq}.
At $\mud=0.007$ the kurtosis of the order parameter starts to overshoot the value $3$ for $\beta\lesssim\betaC$, which is due
to the coexistence of three states and thus clearly signals entering the first-order region.

A few points in \cref{fig:NuMass} were accepted to be obtained from two spatial volumes only.
For the two smallest quark masses which were simulated, it was clear from the crossing point of the kurtosis on $\Ns\in\{12,18,24\}$ that $\Ns=12$ was too far away from the thermodynamic limit.
On the other hand, to add a larger spatial extent would have been very costly without the guarantee to be sufficient for a conclusive statement.
About $\mud=0.75$, instead, we considered the outcome of the analysis with $\Ns\in\{36,42\}$ satisfactory, since the kurtosis of the order parameter reaches values larger than $3.5$ for $\beta\lesssim\betaC$ and the bump shrinks and get larger increasing $\Ns$, behavior typical of the first-order region.

After these considerations, our estimates of the tricritical bare quark masses are
\begin{equation}\label{eq:bareTric}
    \begin{aligned}
        \mTricHat[\text{light}]&=0.007^{+0.002}_{-0.003}\\
        \mTricHat[\text{heavy}]&=\num{0.55(10)}
    \end{aligned}
    \quad,
\end{equation}
where the conservative choice of having an asymmetric error in the chiral region is to stress that further investigation would be needed in the chiral limit to polish the measurement.

\begin{figure*}
    \centering
    %Table
    \begingroup
        \renewcommand{\arraystretch}{1.2}
        \centering
        \small
        \begin{tabular}{*{3}{c@{\hspace{6mm}}}*{3}{S[table-format=1.4(2)]@{\hspace{6mm}}}c@{\hspace{6mm}}c}
              \toprule
              $\Nt$ & $\mud$ & $\beta$ & {$w_0/a$} & {$a\,\mpi$} &
              {$a\!$ \{\si{\femto\metre}\}} & $\mpi$ \{\si{\mega\electronvolt}\} &
              $T\!$ \{\si{\mega\electronvolt}\} \\
              \midrule
              \multirow{3}{*}{4}
              & 0.038 & 5.356 & 0.75505(26) & 0.5236(7) & 0.2324(24) & 445(5) & 212.2(2.2) \\
              & 0.043 & 5.362 & 0.75917(24) & 0.5540(6) & 0.2312(24) & 473(5) & 213.4(2.2) \\
              & 0.048 & 5.368 & 0.76356(22) & 0.5846(7) & 0.2298(24) & 502(5) & 214.6(2.3) \\
              \midrule
              \multirow{8}{*}{6}
              & 0.004 & 5.4324 & 1.1542(9)  & 0.190(4)   & 0.1521(16) &  247(5) & 216.3(2.3) \\
              & 0.005 & 5.4365 & 1.1638(8)  & 0.210(4)   & 0.1509(16) &  275(6) & 218.0(2.3) \\
              & 0.006 & 5.4392 & 1.1643(9)  & 0.238(3)   & 0.1507(16) &  305(5) & 218.2(2.3) \\
              & 0.007 & 5.4425 & 1.1713(9)  & 0.2487(22) & 0.1498(16) &  328(4) & 219.5(2.3) \\
              & 0.008 & 5.4451 & 1.1734(9)  & 0.2651(18) & 0.1496(16) &  350(4) & 219.9(2.3) \\
              & 0.009 & 5.4483 & 1.1799(8)  & 0.2802(18) & 0.1487(16) &  372(5) & 221.1(2.3) \\
              & 0.010 & 5.4515 & 1.1820(8)  & 0.2963(18) & 0.1485(16) &  394(5) & 221.5(2.3) \\
              & 0.011 & 5.4535 & 1.1830(9)  & 0.3066(16) & 0.1483(16) &  408(5) & 221.7(2.3) \\
              \midrule[0.2pt] 
              \multirow{11}{*}{6}
              & 0.150 & 5.6479 & 1.3447(11) & 0.9758(3)  & 0.1305(14) & 1475(16) & 252.0(2.7) \\
              & 0.250 & 5.7118 & 1.3821(11) & 1.2198(3)  & 0.1270(13) & 1896(20) & 259.0(2.7) \\
              & 0.350 & 5.7555 & 1.4118(15) & 1.4136(3)  & 0.1243(13) & 2244(24) & 264.6(2.8) \\
              & 0.400 & 5.7736 & 1.4236(10) & 1.4995(3)  & 0.1233(13) & 2400(25) & 266.8(2.8) \\
              & 0.450 & 5.7878 & 1.4379(11) & 1.5788(3)  & 0.1221(13) & 2552(27) & 269.5(2.8) \\
              & 0.500 & 5.8004 & 1.4422(11) & 1.6544(3)  & 0.1217(13) & 2683(28) & 270.3(2.8) \\
              %\rowcolor{Gray!20}[4mm]
              & 0.550 & 5.8109 & 1.4493(14) & 1.7260(3)  & 0.1211(13) & 2813(30) & 271.6(2.9) \\
              & 0.600 & 5.8201 & 1.4543(18) & 1.7938(3)  & 0.1207(13) & 2933(31) & 272.5(2.9) \\
              & 0.650 & 5.8279 & 1.4583(23) & 1.8592(3)  & 0.1203(13) & 3048(32) & 273.3(2.9) \\
              & 0.750 & 5.8411 & 1.4636(14) & 1.9830(3)  & 0.1199(13) & 3263(34) & 274.3(2.9) \\
              & 0.850 & 5.8512 & 1.4708(13) & 2.0992(4)  & 0.1193(13) & 3471(37) & 275.6(2.9) \\
              \bottomrule 
        \end{tabular}
        \captionof{table}{Results of the scale setting.
                 $T=0$ simulations have been performed on $\protect\lattice{32}{16}$ lattices always collecting $800$ independent configurations.
                 $w_0/a$ has been determined and converted to physical scales using the publicly available code described in~\cite{Borsanyi:2012zs}.
                 For the pion mass determination, $8$ point sources per configuration have been used.
                 The table also contains the lattice spacing, the pion mass and the temperature of the corresponding finite temperature ensemble in physical units.}
        \label{tab:scaleSetting}
    \endgroup
    %Figure
    \vspace{2mm}
    \includegraphics[width=0.9\textwidth, clip, trim=0 2mm 0 0]{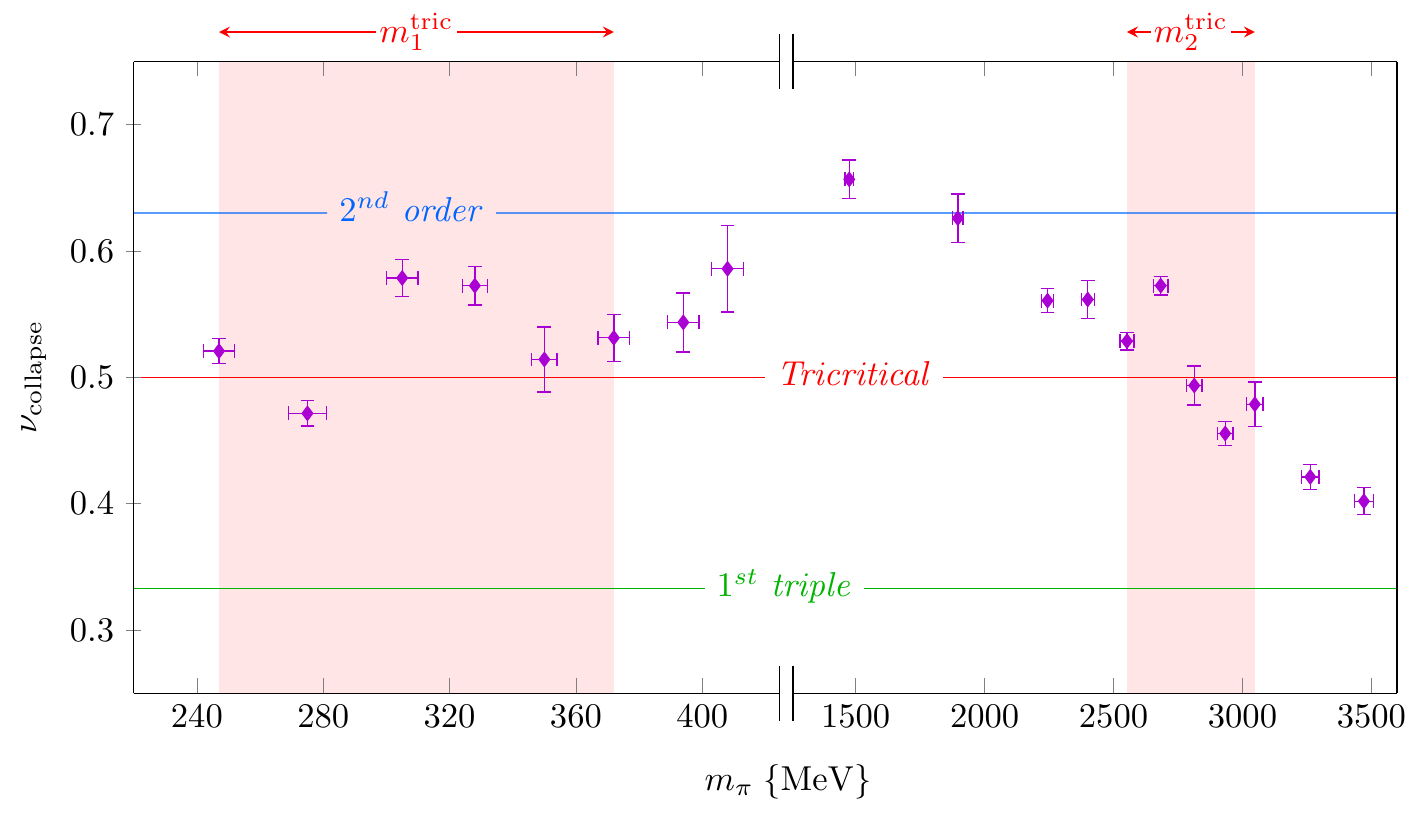}
    \caption{Critical exponent $\nu$ as function of the pion mass $\mpi$.
             This plot is similar to that in \cref{fig:NuMass}, but the values of the critical exponent are only those obtained with the strategy described in \cref{sec:collapse}.
             The regions with light-red background denote the position of the tricritical masses within one standard deviation.}
    \label{fig:NuMpi}
\end{figure*}

In order to asses how much the results are affected by cut-off effects, we measured both the lattice spacing $a$ and the pion mass $\mpi$ for all simulated bare quark masses, by running $T=\MuIpu=0$ simulations at the $\betaC$ found in the \RW\ plane.
The outcome is reported in \cref{tab:scaleSetting}.
Having fixed the scale, it is possible to express \cref{eq:bareTric} in terms of pion masses in physical units,
\begin{equation}\label{eq:physTric}
    \begin{aligned}
        \mTric[\pi,\,\text{light}]&=328^{+44}_{-81}\;\si{\mega\electronvolt}\\
        \mTric[\pi,\,\text{heavy}]&=2813^{+235}_{-261}\;\si{\mega\electronvolt}
    \end{aligned}
    \quad.
\end{equation}
Our results in physical units are given in \cref{fig:NuMpi}, where the critical exponent $\nu$ obtained with our new analysis strategy 
is plotted as a function of $\mpi$.
It is important to stress that in the large-mass region the lattices used are still too coarse to correctly resolve the pion and we 
have $a\mpi>1$, implying sizeable cut-off effects on this value. 

We now compare with the previous results obtained on $\Nt=4$ lattices~\cite{Bonati:2010gi}.
However, there, only the tricritical bare quark masses and a rough estimate of $\mTric[\pi,\,\text{light}]$ were reported.
We therefore improve the determination of the latter, by performing additional scale setting simulations, whose outcome can be found in \cref{tab:scaleSetting}.
In particular, we measured $\mpi$ and $a$ for three values of the quark bare mass, corresponding to the light tricritical point quoted in \refcite{Bonati:2010gi} (the central value and at one standard deviation apart from it).
The value of $\beta$ has been chosen using a polynomial interpolation of the $\betaC$ obtained by the authors at the simulated masses.
Taking as error on the tricritical pion mass the difference between its value and $\mpi$ resulting from the neighboring bare masses, 
we obtain
\begin{equation}
    m_{\pi,\,\text{light}}^{\text{tric},\,\Nt=4}=473^{+29}_{-28}\;\si{\mega\electronvolt}\;.
\end{equation}
We thus conclude that, in the light region, a shift of around $44$\% is found when moving from a $\Nt=4$ to a finer 
$\Nt=6$ lattice.
In the heavy mass region, only a rough comparison is possible, since no pion mass is reported in~\refcite{Bonati:2010gi} and in any case 
on $\Nt=4$ the pion is resolved even less.
However, it is possible to compare the dimensionless ratio $\mud/T$ at the tricritical point,
\begin{equation}
%     \left\{
    \begin{aligned}
        &\\[-2ex]
        \frac{m_{\pi,\,\text{heavy}}^{\text{tric},\,\Nt=4}}{T}&=\num{2.9(3)}\\[1.8ex] % \quad\text{and}\quad
        \frac{m_{\pi,\,\text{heavy}}^{\text{tric},\,\Nt=6}}{T}&=\num{3.3(6)}\\[0.8ex] %\;,
    \end{aligned}
    \quad ,
\end{equation}
which turn out to be compatible.

%================================================================%
\section{Discussion and conclusions}\label{sec:summary}

\begin{figure*}
    \centering
    \includegraphics[width=\textwidth]{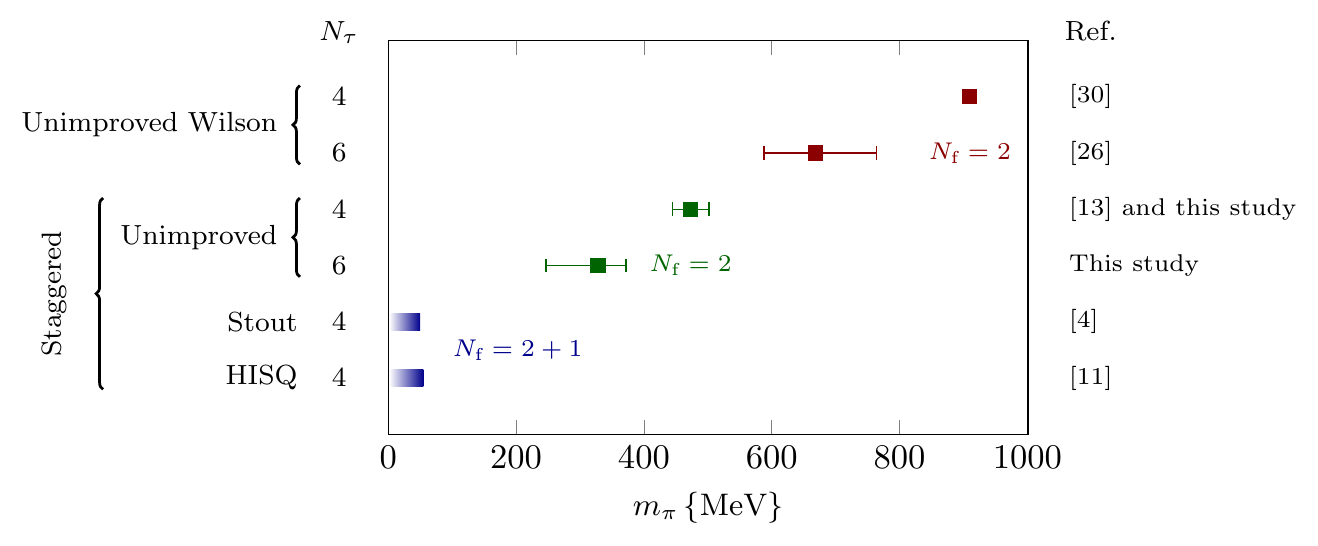}
    \caption{%
        Overview of chiral tricritical values of the pion mass in the \RW\ plane.
    }
    \label{fig:ComparisonTricPoints}
\end{figure*}

Since we are still far from being able to perform a continuum extrapolation, it is instructive to compare 
with other discretizations.
The results obtained in \refcite{Cuteri:2015qkq} with Wilson fermions on $\Nt=6$ lattices, i.e.~with similar lattice spacing,
appear to have considerably larger cut-off effects.
For example, comparing \mbox{$a\mTric[\pi,\,\text{heavy}]=\num{2.2302(2)}$} from  \refcite{Cuteri:2015qkq} with our \mbox{$a\mTric[\pi,\,\text{heavy}]=\num{1.7260(3)}$}, the pion-resolution problem is milder in the present study.
It is also interesting to compare the position of the tricritical points in physical units,
\begin{equation}\label{eq:bareTricChiral}
    \begin{aligned}
        m_{\pi,\,\text{light}}^{\text{tric, Wilson}}&=669^{+95}_{-81}\;\si{\mega\electronvolt}\\
        m_{\pi,\,\text{light}}^{\text{tric, Staggered}}&=328^{+44}_{-81}\;\si{\mega\electronvolt}
    \end{aligned}
\end{equation}
and
\begin{equation}\label{eq:bareTricHeavy}
    \begin{aligned}
        m_{\pi,\,\text{heavy}}^{\text{tric,Wilson}}&=3659^{+589}_{-619}\;\si{\mega\electronvolt}\\
        m_{\pi,\,\text{heavy}}^{\text{tric, Staggered}}&=2813^{+235}_{-261}\;\si{\mega\electronvolt}
    \end{aligned}
    \quad.
\end{equation}
The large differences between discretizations again imply being far from the continuum limit,
where results from all discretizations have to merge. The observed trend is consistent with the findings of
simulations with improved staggered actions, where the tricritical points can only be bounded to be at
much smaller masses, as indicated in \cref{fig:ComparisonTricPoints}, 
as well as with the analogous findings at zero chemical potential (see discussion in the introduction).
In particular the comparison across discretizations implies 
enormous cut-off effects in the critical masses, which could end up being over $\sim$100\% of an
eventual continuum limit. 
We remark that cut-off effects in the critical temperatures are much milder. At present, there is no theoretical
explanation as to why the discretization effects on critical quark masses in the Columbia plot are so strong.

In conclusion, we have determined the shift of the tricritical points in the Roberge-Weiss plane of unimproved staggered fermions
by changing from $\Nt=4$ to $\Nt=6$ lattices. 
The aspect ratios and statistics required to extract the correct order of the phase transition are found to be larger in the 
Roberge-Weiss plane than at $\mu=0$.
We find the cut-off effect on the tricritical masses to be smaller but qualitatively the same as that observed with Wilson fermions,
and consistent with results for both discretizations at zero chemical potential. 
This implies in particular, that the entire chiral critical surface depicted
in \cref{fig:CP3dI} is shifted significantly towards smaller (and possibly zero) light quark masses, as the lattice spacing decreases,
which is also consistent with results from improved staggered actions. 
Unfortunately, our study also implies that much finer lattices at inevitably smaller quark masses are necessary, 
before one can hope the results of the light tricritical mass to stabilize in a continuum limit.

%================================================================%
\begin{acknowledgments}
    We thank Francesca Cuteri for useful discussions and input for \cref{fig:ComparisonTricPoints}.
    The authors acknowledge support by the Deutsche Forschungsgemeinschaft (DFG) through the grant CRC-TR 211 ``Strong-interaction matter
    under extreme conditions'' and by the Helmholtz International Center for FAIR within the LOEWE program of the State of Hesse.
    We also thank the computing staff of \Lcsc\ for their support. 
\end{acknowledgments}
\vspace*{35mm}
%================================================================%
\bibliographystyle{apsrev4-1}
\bibliography{literature.bib}
%================================================================%
\onecolumngrid
\appendix*

\section{Simulation details}\label{app:sim}

It is known that different powers of the same observable have different integrated autocorrelation times $\tauInt$, which can be estimated using the Wolff algorithm~\cite{Wolff:2003sm}.
This is important to be taken into account when it comes to measure standardized cumulants, like the kurtosis of a given observable.
Binning, i.e. substituting a block of data with its average, allows to obtain uncorrelated data from the correlated ones.
This is true if the size of a block is at least twice $\tauInt$, though.
It is then possible to understand how many independent measurement of the quantity of interest are available in a Monte Carlo simulation, just by dividing the number of the trajectories produced by $2\tauInt$.
Clearly, the larger this number is the more accurate the result will be.
However, simulations in full \qcd\ are costly and a compromise is needed.
We always had at least $100$ independent events for $\kurtosis(\PolyIm)$ in the merged chain obtained by putting together the four independent Markov chain that we produced for each $\beta$ value.
A detailed overview of the collected statistics is presented in \cref{tab:statistics}.

\Cref{tab:collapse,tab:fits} contain, instead, the detailed outcome of our analysis, whose data were plotted in \cref{fig:NuMass}.

\begingroup
    % Command to get entries of the table (from where mass folders are)
    %
    % echo; for m in mass*; do cd $m/nt6; for ns in ns*; do cd $ns; echo "$m $ns eventsMin=$(grep -v "merged"
    % Nf2_muiPiT_${m}_nt6_${ns}_analysis/Nf2_muiPiT_${m}_nt6_${ns}_observables_poly_im_withZeroMean.dat
    % | awk 'NR==2{sum=$25; totStat=$3; min=$28}NR>2{sum+=$25; totStat+=$3; if($28<min){min=$28}}
    % END{printf "%d\ttauIntMean=%.0f\t\ttotal statistics=%.2fM", min, sum/(NR-1), totStat/1.e6}')   
    % $(awk 'NR==1{printf "betas = [%s", $1}END{printf " - %s]", $1}' betas)""; cd ../; done; cd ../../; done; echo
    \newcommand{\CC}[1][gray]{\cellcolor{#1}}
    \renewcommand{\arraystretch}{1.5}
    \def\colorXII{cyan!20}
    \def\colorXVIII{magenta!20}
    \def\colorXXXVI{yellow!20}
    \def\colorXLII{orange!20}
    \renewcommand{\c}[1]{\tc{red}{#1}}
    \begin{table}[ht]
        \centering
        \setlength{\aboverulesep}{0pt}
        \setlength{\belowrulesep}{0pt}
        \setlength{\tabcolsep}{3pt}
        \scriptsize
        \begin{tabular}{*{6}{c}}
            \toprule
            \multirow{2}{*}{$\mud$} & \multirow{2}{*}{$\beta$ range} &
            \multicolumn{4}{c}{Total statistics per spatial lattice size $\Ns$ $\bigl($ \texttt{\#} of simulated 
            $\;\beta\mid\bar{\tau}_{\text{\textsc{int}}}\mid n_{\text{min}}^{\text{events}}$ per chain $\bigr)$ } \\
            & & \colorbox{\colorXII}{12} \colorbox{\colorXXXVI}{36} & \colorbox{\colorXVIII}{18} \colorbox{\colorXLII}{42} & 24 & 30 \\
            \midrule
            0.004 & 5.425~-~5.437 & \CC[\colorXII]   ~1.56M~(5~|~~~88~|~301) & \CC[\colorXVIII]~1.56M~(4~|~~247~|~118) & ~1.32M~(4~|~339~|~~64) &            -            \\
            0.005 & 5.427~-~5.442 & \CC[\colorXII]   ~1.00M~(5~|~~~86~|~225) & \CC[\colorXVIII]~3.04M~(5~|~~199~|~153) & ~2.12M~(5~|~310~|~~63) &            -            \\
            0.006 & 5.430~-~5.445 & \CC[\colorXII]   ~2.28M~(6~|~~~91~|~363) & \CC[\colorXVIII]~1.92M~(6~|~~212~|~~84) & ~1.08M~(4~|~410~|~~61) & ~1.00M~(4~|~~527~|~~23) \\
            0.007 & 5.420~-~5.460 & \CC[\colorXII]   ~0.64M~(5~|~~~69~|~174) & \CC[\colorXVIII]~1.72M~(5~|~~174~|~135) & ~1.10M~(5~|~256~|~~63) & ~1.56M~(4~|~~420~|~~39) \\
            0.008 & 5.430~-~5.470 & \CC[\colorXII]   ~0.92M~(5~|~~~66~|~217) & \CC[\colorXVIII]~0.86M~(5~|~~175~|~~68) & ~1.46M~(4~|~331~|~~57) &            -            \\
            0.009 & 5.430~-~5.470 & \CC[\colorXII]   ~1.26M~(5~|~~~72~|~266) & \CC[\colorXVIII]~1.28M~(5~|~~182~|~~91) & ~1.60M~(4~|~325~|~~66) &            -            \\
            0.010 & 5.430~-~5.480 & \CC[\colorXII]   ~1.20M~(6~|~~~65~|~297) & \CC[\colorXVIII]~0.60M~(4~|~~143~|~~71) & ~1.84M~(5~|~263~|~~48) &            -            \\
            0.011 & 5.430~-~5.490 & \CC[\colorXII]   ~0.68M~(4~|~~~57~|~229) & \CC[\colorXVIII]~1.24M~(5~|~~180~|~~82) & ~1.92M~(4~|~336~|~~72) &            -            \\
            \midrule[0.2pt]
            0.150 & 5.590~-~5.720 & \CC[\colorXII]   ~1.08M~(5~|~~~62~|~275) & \CC[\colorXVIII]~5.80M~(7~|~~242~|~203) & ~5.28M~(7~|~362~|~153) &            -            \\
            0.250 & 5.600~-~5.760 & \CC[\colorXII]   ~4.20M~(7~|~~~81~|~555) & \CC[\colorXVIII]~2.00M~(4~|~~190~|~167) & ~4.60M~(6~|~409~|~~85) &            -            \\
            0.350 & 5.720~-~5.780 & \CC[\colorXII]   ~6.44M~(7~|~~130~|~510) & \CC[\colorXVIII]~3.40M~(5~|~~279~|~131) & ~5.88M~(6~|~442~|~130) &            -            \\
            0.400 & 5.750~-~5.790 &                    -                     & \CC[\colorXVIII]~9.00M~(5~|~~305~|~214) & 10.60M~(6~|~574~|~116) & ~7.80M~(5~|~~917~|~~95) \\
            0.450 & 5.760~-~5.810 & \CC[\colorXXXVI] 18.40M~(4~|~1343~|~287) & \CC[\colorXVIII]~4.72M~(5~|~~330~|~189) & 20.00M~(5~|~636~|~511) & 20.00M~(5~|~1010~|~279) \\
            0.500 & 5.780~-~5.820 & \CC[\colorXXXVI] 10.00M~(4~|~1507~|~107) & \CC[\colorXVIII]~9.40M~(5~|~~325~|~255) & 10.00M~(5~|~598~|~112) & 11.20M~(5~|~~889~|~236) \\
            0.550 & 5.760~-~5.840 & \CC[\colorXXXVI] ~6.20M~(5~|~1355~|~~38) & \CC[\colorXVIII]10.80M~(6~|~~300~|~173) & ~6.56M~(5~|~606~|~~77) & ~7.80M~(5~|~~917~|~~57) \\
            0.600 & 5.812~-~5.827 & \CC[\colorXXXVI] 12.00M~(4~|~1720~|~152) &                    -                    & 19.80M~(6~|~766~|~404) & 14.00M~(6~|~1207~|~~94) \\
            0.650 & 5.817~-~5.837 & \CC[\colorXXXVI] 15.40M~(4~|~1759~|~~74) & \CC[\colorXVIII]16.40M~(5~|~~501~|~673) & 20.40M~(5~|~780~|~295) & 15.40M~(5~|~1301~|~192) \\
            0.750 & 5.830~-~5.846 & \CC[\colorXXXVI] 15.20M~(4~|~1976~|~~79) & \CC[\colorXLII] ~5.60M~(3~|~2507~|~~48) &            -           & 16.80M~(5~|~1240~|~~93) \\
            0.850 & 5.840~-~5.856 & \CC[\colorXXXVI] 19.60M~(4~|~1792~|~111) & \CC[\colorXLII] 18.00M~(5~|~2066~|~~71) &            -           & 16.40M~(4~|~1288~|~136) \\
            \bottomrule
        \end{tabular}
        \normalsize
        \caption{Overview of the statistics accumulated in all the simulations (red entries are preliminary).
                 Since the resolution in $\beta$ is not the same at different $\mud$, the number of simulated
                 $\beta$ has been reported per each range.
                 The accumulated statistics per $\beta$ varies because of the criterion adopted to stop to increase the
                 statistics on the $4$ chains.
                 Therefore we reported here the total number of trajectories produced per given $\Ns$.
                 For each $\Ns$, the number of simulated $\beta$, the average integrated autocorrelation time and the
                 smallest number of independent events per chain of $\kurtosis(\PolyIm)$ can be found in the brackets
                 next to the total statistics.
                 Observe that $\bar{\tau}_{\text{\textsc{int}}}$ and $n_{\text{min}}^{\text{events}}$ are not connected.
                 The former is an average among all the different chains run at one fixed spatial lattice extent, while
                 the latter is the effective length of the shorter chain for that given $\Ns$.
                 The number of independent events is obtained as ratio between the number of produced trajectories and the
                 bin size, which is roughly $2\,\tauInt$.}
        \label{tab:statistics}
    \end{table}
\endgroup
\begingroup
    \renewcommand{\arraystretch}{1.4}
    \renewcommand{\b}[1]{\tc{blue}{#1}}
    \renewcommand{\g}[1]{\tc{ForestGreen}{#1}}
    \renewcommand{\r}[1]{\tc{Red}{#1}}
    \renewcommand{\c}[1]{\tc{Cyan}{#1}}
    \newcommand{\p}[1]{\tc{purple}{#1}}
    \renewcommand{\o}[1]{\tc{Gray!50!black}{#1}}
    \newcolumntype{L}{>{\arraybackslash}c}

    \begin{table}
        \setlength{\aboverulesep}{0pt}
        \setlength{\belowrulesep}{0pt}
        \centering
        \small
        \sisetup{table-number-alignment = center}
        {
            \begin{tabular}{*{2}{c@{\hspace{3mm}}}S[table-format = 1.6(2)]@{\hspace{3mm}}S[table-format = 1.3(2)]}
                \toprule
                $\mud$ & $\Ns$ & $\betaC^{\text{extr.}}$ & $\nu^{\text{extr.}}$ \\[0.5mm]
                \midrule
                0.004 &        \g{18} \r{24} & 5.43261 (5)  & 0.521(10) \\
                0.005 &        \g{18} \r{24} & 5.43648 (4)  & 0.471(10) \\
                0.006 & \g{18} \r{24} \c{30} & 5.43917 (5)  & 0.579(15) \\
                0.007 & \g{18} \r{24} \c{30} & 5.44260 (5)  & 0.573(15) \\
                0.008 & \b{12} \g{18} \r{24} & 5.44529 (8)  & 0.514(26) \\
                0.009 & \b{12} \g{18} \r{24} & 5.44834 (10) & 0.531(19) \\
                0.010 & \b{12} \g{18} \r{24} & 5.45168 (12) & 0.544(23) \\
                0.011 & \b{12} \g{18} \r{24} & 5.453888(11) & 0.59 (3)  \\
                \bottomrule
            \end{tabular}
        } \qquad
        {
            \begin{tabular}{*{2}{c@{\hspace{3mm}}}S[table-format = 1.5(2)]@{\hspace{3mm}}S[table-format = 1.3(2)]}
                \toprule
                $\mud$ & $\Ns$ & $\betaC^{\text{extr.}}$ & $\nu^{\text{extr.}}$ \\[0.5mm]
                \midrule
                0.150 &        \b{12} \g{18} \r{24} & 5.64761(15) & 0.657(15)  \\
                0.250 &        \b{12} \g{18} \r{24} & 5.71218(5)  & 0.626(19)  \\
                0.350 &        \b{12} \g{18} \r{24} & 5.75559(11) & 0.561(10)  \\
                0.400 &        \g{18} \r{24} \c{30} & 5.77373(7)  & 0.562(15)  \\
                0.450 & \g{18} \r{24} \c{30} \p{36} & 5.78783(8)  & 0.529(7)   \\
                0.500 &        \r{24} \c{30} \p{36} & 5.80070(4)  & 0.573(7)   \\
                0.550 &        \r{24} \c{30} \p{36} & 5.81096(6)  & 0.494(16)  \\
                0.600 &        \r{24} \c{30} \p{36} & 5.81999(5)  & 0.456(10)  \\
                0.650 &        \r{24} \c{30} \p{36} & 5.82777(5)  & 0.479(18)  \\
                0.750 &               \p{36} \o{42} & 5.84114(5)  & 0.421(10)  \\
                0.850 &        \c{30} \p{36} \o{42} & 5.85135(4)  & 0.402(11)  \\
                \bottomrule
            \end{tabular}
        }
        \caption{Result of the quantitative collapse analysis.
                 The critical temperature $\betaC$ and the critical exponent $\nu$ have been found minimizing $Q(\bar{\betaC},\bar{\nu})$ as defined in \cref{eq:qualityCollapse} for several decreasing values of $\Delta x$.
                 $\betaC^{\text{extr.}}$ and $\nu^{\text{extr.}}$ are the outcome of a linear extrapolation for $\Delta x\to0$.
                 Note that the reweighting resolution in $\beta$ used to add new points between simulated ones varied between $0.004$ and $0.0002$ and it has been chosen in order to have around $20$ values of the kurtosis to be later interpolated.}
        \label{tab:collapse}
    \end{table}
\endgroup

\clearpage
\newpage
\global\pdfpageattr\expandafter{\the\pdfpageattr/Rotate 90}
%----------------------------------------------------------------%
\begingroup
    % Command to get fitted points
    %
    % Two fitted volumes
    % awk '$1 ~ /^(#|$)/{next}{if(split($2, vols, ".")==2){print $0}}' Nt6_fromFit.dat | 
    % awk '{print $1, $16, $17, $18, $19, $20, $3}' | sed 's/_/ /g' | awk '{print $1, $6, " -> ", ($3-$2)/$7+1, $8, " -> ", ($5-$4)/$9+1, $12}'
    % 
    % Three fitted volumes
    % awk '$1 ~ /^(#|$)/{next}{if(split($2, vols, ".")==3){print $0}}' Nt6_fromFit.dat | 
    % awk '{print $1, $16, $17, $18, $19, $20, $21, $22, $3}' | sed 's/_/ /g' | awk '{print $1, $8, " -> ", ($3-$2)/$9+1, $10, " -> ", ($5-$4)/$11+1, $12, " -> ", ($7-$6)/$13+1, "    ", $14}'
    
    \renewcommand{\arraystretch}{1.5}
    \renewcommand{\b}[1]{\tc{blue}{#1}}
    \renewcommand{\g}[1]{\tc{ForestGreen}{#1}}
    \renewcommand{\r}[1]{\tc{Red}{#1}}
    \renewcommand{\c}[1]{\tc{Cyan}{#1}}
    \newcommand{\p}[1]{\tc{purple}{#1}}
    \renewcommand{\o}[1]{\tc{Gray!50!black}{#1}}
    \newcommand{\mpb}[1]{\parbox[b]{\widthof{0.5}}{\centering #1}}
    \newcommand{\sd}[1]{\phantom{1}#1} % ReVTeX 4-1

    \newcolumntype{L}{>{\arraybackslash}r}
    \newcolumntype{G}{>{\columncolor[gray]{0.2}[.5\tabcolsep][\tabcolsep]}c}
    \newcolumntype{I}{>{\columncolor[gray]{0.2}[\tabcolsep][.5\tabcolsep]}c}

    \begin{sidewaystable}
        \setlength{\aboverulesep}{0pt}
        \setlength{\belowrulesep}{0pt}
        \centering
        \sisetup{table-number-alignment = center}
        \begin{tabular}{@{\hspace{4mm}}*{3}{r@{\hspace{6mm}}}L@{\hspace{8mm}}S[table-format = 1.5(2)]*{2}{@{\hspace{4mm}}S[table-format = 1.3(3)]}@{\hspace{4mm}}S[table-format = +1.3(2)]@{\hspace{4mm}}*{4}{c@{\hspace{4mm}}}}
            \toprule
            $\mud$ & $\Ns$ & \texttt{\#} points & $\delta\beta\cdot10^{3}$ & {$\betaC$} & {$\nu$} &
            {$\kurtosis(\betaC,\infty)$} & {$a_1$} & $\chi^2_{_{\text{d.o.f.}}}$ & Q(\%) & $\Omega_{\text{min}}$ & $\Xi_{\text{min}}$ \\[0.5mm]
            \midrule
            0.004 &         \g{18}~~\r{24}         & \g{24}~~\r{\sd{8}}                        &            \g{0.3}~~\r{0.2}     & 5.43241(10) & 0.52  (8)  & 1.794(14) & -0.44(40)  & 0.864 & 67.18 & 52.21 & 72.17 \\
            0.005 &         \g{18}~~\r{24}         & \g{11}~~\r{\sd{7}}                        &            \g{0.5}~~\r{0.5}     & 5.43649(11) & 0.49  (4)  & 1.699(14) & -0.27(14)  & 1.004 & 44.56 & 96.06 & 99.33 \\
            0.006 &         \g{18}~~\r{24}~~\c{30} & \g{8}~~\r{\sd{5}}~~\c{13}                 & \g{1}~~\r{\mpb{1}}~~\c{0.3}     & 5.43918(7)  & 0.554(26)  & 1.823(11) & -0.55(14)  & 1.005 & 45.27 & 84.11 & 78.89 \\
            0.007 &         \g{18}~~\r{24}~~\c{30} & \g{5}~~\r{\sd{4}}~~\c{\sd{3}}             & \g{1}~~\r{0.5}~~\c{0.5}         & 5.44247(10) & 0.48  (6)  & 1.780(14) & -0.21(17)  & 0.768 & 63.12 & 81.44 & 62.67 \\
            0.008 &         \b{12}~~\g{18}~~\r{24} & \b{7}~~\g{\sd{6}}~~\r{\sd{6}}             & \b{2}~~\g{\mpb{1}}~~\r{0.5}     & 5.44513(11) & 0.472 (26) & 1.878(10) & -0.23(7)   & 0.995 & 45.70 & 94.98 & 85.20 \\
            0.009 &         \b{12}~~\g{18}~~\r{24} & \b{6}~~\g{\sd{8}}~~\r{\sd{9}}             & \b{2}~~\g{\mpb{1}}~~\r{0.5}     & 5.44828(10) & 0.540 (25) & 1.871(10) & -0.44(11)  & 1.001 & 45.51 & 80.52 & 85.60 \\
            0.010 &         \b{12}~~\g{18}~~\r{24} & \b{6}~~\g{\sd{5}}~~\r{\sd{8}}             & \b{3}~~\g{\mpb{2}}~~\r{\mpb{1}} & 5.45154(14) & 0.623 (26) & 1.894(12) & -0.88(16)  & 1.005 & 44.65 & 80.74 & 72.80 \\
            0.011 &         \b{12}~~\g{18}~~\r{24} & \b{6}~~\g{\sd{8}}~~\r{11}                 & \b{4}~~\g{\mpb{2}}~~\r{\mpb{1}} & 5.45350(16) & 0.650 (27) & 1.999(18) & -1.07(20)  & 0.994 & 46.71 & 81.53 & 64.29 \\
            \midrule[0.2pt]
            0.150 &         \b{12}~~\g{18}~~\r{24} & \b{6}~~\g{16}~~\r{10}                     & \b{4}~~\g{\mpb{1}}~~\r{\mpb{1}} & 5.64787(23) & 0.66  (4)  & 1.854(13) & -0.65(18)  & 1.007 & 45.43 & 83.79 & 68.40 \\
            0.250 &         \b{12}~~\g{18}~~\r{24} & \b{8}~~\g{23}~~\r{11}                     & \b{4}~~\g{\mpb{1}}~~\r{\mpb{1}} & 5.71176(22) & 0.652 (26) & 1.914(10) & -0.57(10)  & 0.990 & 48.67 & 81.19 & 87.43 \\
            0.350 &         \b{12}~~\g{18}~~\r{24} & \b{5}~~\g{\sd{8}}~~\r{10}                 & \b{5}~~\g{\mpb{2}}~~\r{\mpb{1}} & 5.75551(16) & 0.596 (18) & 1.900(7)  & -0.39(6)   & 0.997 & 46.08 & 81.96 & 78.00 \\
            0.400 &         \g{18}~~\r{24}~~\c{30} & \g{5}~~\r{\sd{7}}~~\c{10}                 & \g{2}~~\r{\mpb{1}}~~\c{0.5}     & 5.77364(14) & 0.59  (4)  & 1.777(11) & -0.36(12)  & 0.992 & 46.57 & 85.59 & 72.89 \\
            0.450 & \g{18}~~\r{24}~~\c{30}~~\p{36} & \g{4}~~\r{\sd{3}}~~\c{\sd{5}}~~\p{\sd{5}} & \g{2}~~\r{2}~~\c{0.5}~~\p{0.5}  & 5.78783(7)  & 0.54  (3)  & 1.777(8)  & -0.23(8)   & 0.997 & 45.11 & 83.15 & 83.00 \\
            0.500 &         \r{24}~~\c{30}~~\p{36} & \r{4}~~\c{\sd{7}}~~\p{\sd{7}}             & \r{2}~~\c{0.5}~~\p{0.5}         & 5.80037(18) & 0.60  (6)  & 1.735(23) & -0.44(23)  & 0.994 & 45.61 & 83.99 & 58.00 \\
            0.550 &         \r{24}~~\c{30}~~\p{36} & \r{4}~~\c{12}~~\p{\sd{8}}                 & \r{2}~~\c{0.5}~~\p{0.5}         & 5.81091(13) & 0.51  (4)  & 1.783(20) & -0.17(10)  & 0.997 & 46.22 & 82.53 & 90.86 \\
            0.600 &         \r{24}~~\c{30}~~\p{36} & \r{6}~~\c{\sd{8}}~~\p{\sd{6}}             & \r{1}~~\c{0.5}~~\p{0.5}         & 5.82012(6)  & 0.457 (24) & 1.758(9)  & -0.09(4)   & 0.991 & 46.26 & 90.35 & 70.40 \\
            0.650 &         \r{24}~~\c{30}~~\p{36} & \r{7}~~\c{\sd{8}}~~\p{\sd{8}}             & \r{1}~~\c{0.5}~~\p{0.5}         & 5.82785(9)  & 0.481 (28) & 1.813(12) & -0.13(5)   & 0.991 & 46.80 & 81.26 & 94.29 \\
            0.750 &                 \p{36}~~\o{42} &                \p{15}~~\o{16}             &        \p{0.3}~~\o{0.2}         & 5.84112(10) & 0.39  (4)  & 1.86(3)   & -0.031(29) & 0.992 & 47.54 & 93.13 & 58.67 \\
            0.850 &         \c{30}~~\p{36}~~\o{42} & \c{5}~~\p{\sd{8}}~~\o{\sd{8}}             & \c{1}~~\p{0.3}~~\o{0.2}         & 5.85124(5)  & 0.358 (25) & 1.807(17) & -0.014(9)  & 0.996 & 45.88 & 83.39 & 51.43 \\
            \bottomrule
        \end{tabular}
        \caption{Overview of the selected fits to extract the final value of $\nu$ (results on grey background are preliminary).
                 The fits have been performed according to \cref{eq:kurtosisExpansion}, considering the linear term only.
                 The $\Ns$ column contains the spatial lattice extents that have been included in the fits.
                 $\Omega_{\text{min}}$ and $\Xi_{\text{min}}$ are respectively the minimum overlap percentage and the minimum symmetry percentage as defined in Eqs.~(B3) and (B4) of \refcite{Cuteri:2015qkq}.
                 In the third and the fourth column, the number of fitted points per $\Ns$ and the reweighting resolution in $\beta$ have been reported, respectively.
                 The colors should help as guideline to distinguish the information among the different volumes.
                 Observe how the smallest volume has systematically a smaller weight in the fit (apart from the two smallest masses, which should anyway investigated more).}
        \label{tab:fits}
    \end{sidewaystable}

    \endgroup
%----------------------------------------------------------------%
\clearpage
\newpage
\global\pdfpageattr\expandafter{\the\pdfpageattr/Rotate 0}

%================================================================%
\end{document}